\documentclass[journal,12pt,onecolumn,draftclsnofoot,]{IEEEtran}
\usepackage[utf8]{inputenc} 
\usepackage[T1]{fontenc}
\usepackage{url}
\usepackage{ifthen}
\usepackage[cmex10]{amsmath}
\usepackage{array} 

\usepackage{enumerate}
\usepackage{amssymb}
\usepackage{amsmath} 

\usepackage{algorithm}
\usepackage{algorithmicx}
\usepackage{algpseudocode}

\usepackage{mathrsfs}
\usepackage{bm}
\usepackage{tikz}
\usetikzlibrary{arrows}
\usepackage{subfigure}
\usepackage{graphicx,booktabs,multirow}
\usepackage{epstopdf}
\usepackage{subfigure}
\usepackage{multirow}
\usepackage{tabularx} 
\usepackage{booktabs}

\usepackage{pifont}

\definecolor{colorhkust}{RGB}{20,43,140}
\definecolor{colortsinghua}{RGB}{116,52,129}
\definecolor{color1}{RGB}{128,0,0}

\usepackage{amsthm}
\usepackage{cite}
\newtheorem{thm}{Theorem}

\newtheorem{prop}{Proposition}

\theoremstyle{definition}

\theoremstyle{remark}

\begin{document}

        \title{Learn to Communicate with Neural Calibration: Scalability and Generalization}

      \author{Yifan Ma, \textit{Graduate Student Member}, \textit{IEEE}, Yifei Shen, \textit{Graduate Student Member}, \textit{IEEE}, Xianghao Yu, \textit{Member}, \textit{IEEE}, Jun Zhang \textit{Senior Member}, \textit{IEEE}, S.H. Song, \textit{Member}, \textit{IEEE}, and Khaled B. Letaief, \textit{Fellow}, \textit{IEEE}
      	\thanks{     		
      	Part of this work has been accepted by IEEE Global Communications Conference, Madrid, Spain, Dec. 2021 \cite{Ma21Neural}. This work was supported by the General Research Fund (Project No. 16210719 and 15207220) from the Hong Kong Research Grants Council. The authors are with the Department of Electronic and Computer Engineering, The Hong Kong University of Science and Technology, Hong Kong (E-mail: \{ymabj, yshenaw, eexyu, eejzhang, eeshsong, eekhaled\}@ust.hk).  (The corresponding author is J. Zhang.)
}
}
        
        \maketitle
        
\begin{abstract}
The conventional design of wireless communication systems typically relies on established mathematical models that capture the characteristics of different communication modules. Unfortunately, such design cannot be easily and directly applied to future wireless networks, which will be characterized by large-scale ultra-dense networks whose design complexity scales exponentially with the network size. Furthermore, such networks will vary dynamically in a significant way, which makes it intractable to develop comprehensive analytical models.
Recently, deep learning-based approaches have emerged as potential alternatives for designing complex and dynamic wireless systems. 
However, existing learning-based methods have limited capabilities to scale with the problem size and to generalize with varying network settings.
In this paper, we propose a scalable and generalizable \emph{neural calibration} framework for future wireless system design, where a neural network is adopted to calibrate the input of conventional model-based algorithms. Specifically, the backbone of a traditional time-efficient algorithm is integrated with deep neural networks to achieve a high computational efficiency, while enjoying enhanced performance. The permutation equivariance property, carried out by the topological structure of wireless systems, is furthermore utilized to develop a generalizable neural network architecture. The proposed neural calibration framework is applied to solve challenging resource management problems in massive multiple-input multiple-output (MIMO) systems. Simulation results will show that the proposed neural calibration approach enjoys significantly improved scalability and generalization compared with the existing learning-based methods.

\end{abstract}

\section{Introduction}
In the past decades, the design of wireless communication systems heavily relied on exploiting domain knowledge based on mathematical models that characterize the signal propagation process and communication system. However, there are cases where the model-based methods meet obstacles, e.g., when the mathematical models are highly complex, can not accurately capture the underlying dynamics of the system, or do not lead to computationally-efficient algorithms \cite{shlezinger21modelbased}. As future wireless networks are expected to experience an unprecedented level of complexity, we are rapidly reaching the point that exceeds the capabilities and applicability of traditional modeling and design approaches \cite{Zappone19Era}. Inspired by the recent success of deep learning in a variety of domains such as computer vision, natural language processing, and recommendation systems \cite{deng14DeepApplication}, deep learning-based algorithms have been proposed to solve the complicated and delay-sensitive tasks for wireless communications. Applications of deep learning in wireless systems have spanned many design problems, including channel estimation\cite{Dong19CNN, He18CE}, data detection \cite{Samuel19Detect, He20Model}, power control \cite{sun2018learning, liang2018towards, lee2018deep}, beamforming design \cite{Lin20Beamform, Shen21Graph, Hu21DeepUnfold}, reflective elements tuning \cite{Jiang20Implicit}, and feature transmission for edge inference \cite{Shao21Edge, shao2021taskoriented}, thriving the emerging research area of ``learning to communicate”.

The targeted high network capacity and ubiquitous wireless connectivity of the fifth-generation (5G) wireless networks have been largely achieved, thanks to key enabling technologies such as ultra-dense networks, massive multiple-input multiple-output (MIMO), millimeter-wave (mm-wave) communication, etc. \cite{5G14}. However, these technologies introduce challenges for learning-based system design methodologies \cite{Letaief196G}. On the one hand, as the network density and the number of antennas increase drastically, the resulting optimization problems become high-dimensional and therefore hard to solve \cite{Shi15Large}. Besides, the channel coherence time is extremely small in higher frequency wireless communication systems \cite{Rangan14mmWave}. Due to fast channel fluctuations, the learning-based algorithms are required to be executed in a time frame of milliseconds \cite{sun2018learning}. The above two challenges require that the deep learning-based methods achieve high \emph{scalability}, namely, to solve large-scale problems with satisfactory performance in real-time. On the other hand, due to the dynamic environment, network settings such as the signal-to-noise ratio (SNR) and the number of served users can change rapidly. Therefore, after being trained on the training dataset, the learning models have to react to the dynamic environment for online inference. 
This ability is referred to as \emph{generalization}, which characterizes the adaptability of the learned models to previously unseen data \cite{goodfellow2016deep}. 
In particular, when the system settings change, a good generalization calls for accurate predictions without re-training or reloading the neural networks. Given the above, it is clear that scalability and generalization are two essential requirements of the ``learning to communicate” approach for future wireless networks.

\begin{figure}[t] 
\centering
\includegraphics[width=0.55\textwidth]{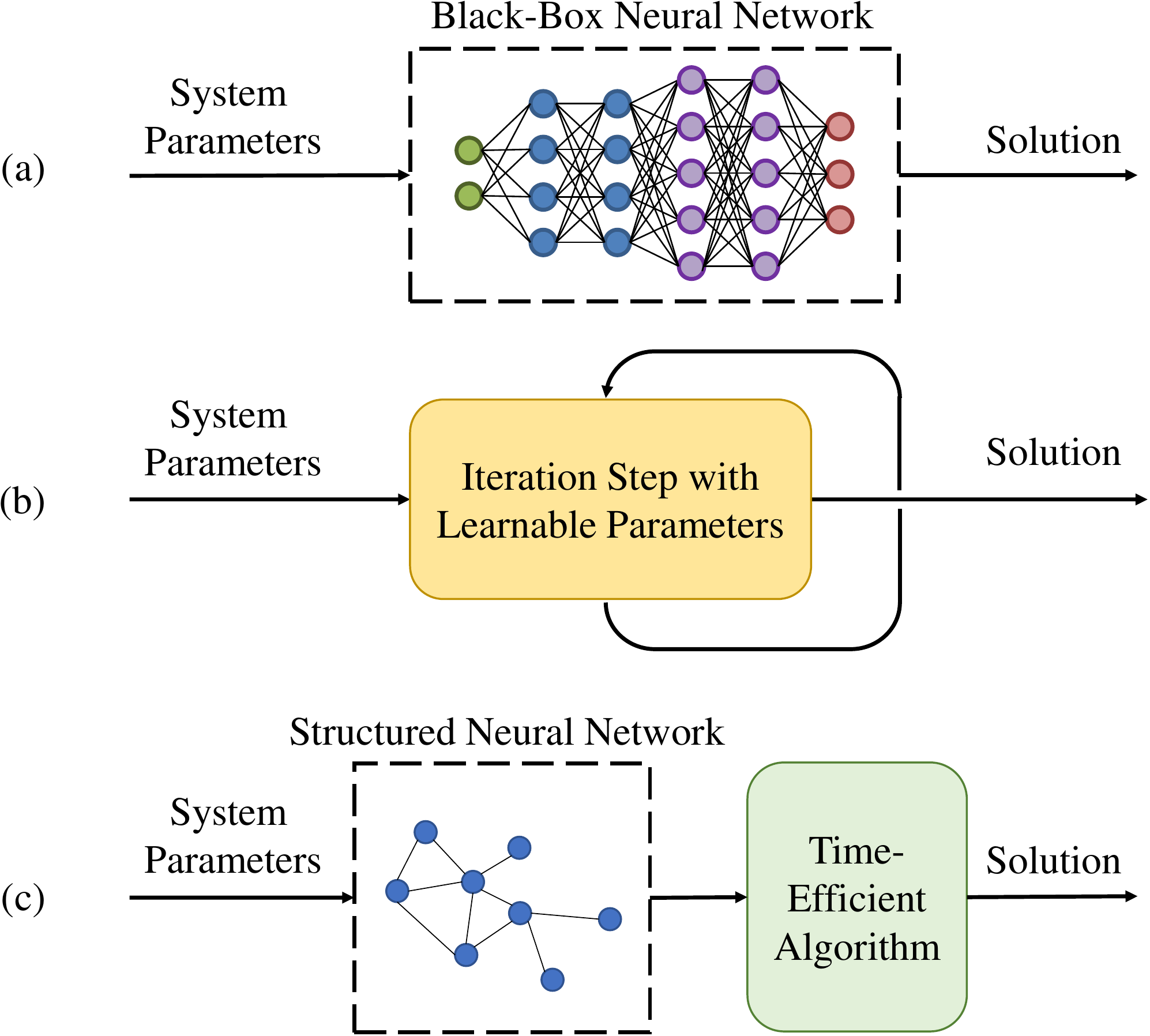} 
\caption{Illustration of different learning-based paradigms: (a) fully data-driven methods; (b) deep unfolding; (c) neural calibration.} 
\label{paradigm} 
\end{figure}

There are two prevalent learning-based paradigms for wireless communication systems \cite{Shen21Graph}. First, most of the existing works focused on \emph{fully data-driven} methods \cite{sun2018learning, liang2018towards,  lee2018deep, Lin20Beamform, Jiang20Implicit, Dong19CNN}, which completely replace conventional model-based algorithms with neural networks to approximate the complicated mapping from the system parameters to the optimal solutions, as shown in Fig. \ref{paradigm}(a). For example, in \cite{sun2018learning}, to tackle the power control problem, the nonlinear mapping from the channel coefficients to the conventional weighted minimum mean-square error (WMMSE) solution was learned using multi-layer perceptron (MLP) and significant speedup was achieved. 
Follow-up works adopted unsupervised losses \cite{liang2018towards} and the convolutional neural network (CNN) \cite{lee2018deep} to achieve a higher spectral efficiency. 
Although fully data-driven methods can be executed in real-time, they generally lack acceptable scalability and generalization ability when the network size becomes large. Specifically, since conventional communication modules are typically treated as black-boxes and replaced by standard neural network architectures, the underlying high-dimensional mapping is difficult to learn in large-scale wireless networks. 
Therefore, the performance of these methods deteriorates dramatically when the number of optimization variables becomes large. For example, the method proposed in \cite{sun2018learning} achieves $98.33\%$ relative sum-rate (relative to WMMSE) when the number of transceivers pairs is $10$, which, however, drops to $84.29\%$ when the number of transceivers pairs grows to $30$. Moreover, since the input and output dimensions of the MLP and CNN are fixed, these methods cannot generalize when the dimension of optimization variables in the test dataset is larger than that in the training dataset \cite{Shen21Graph}. Therefore, the neural networks have to be re-trained once the number of users changes in the wireless networks. One may relieve this issue by training and storing different sizes of neural networks for different network settings, which, however, costs additional computation and memory resources. 

\emph{Deep unfolding} was proposed as another paradigm that combines communication domain knowledge with deep learning to make the learning model explainable \cite{He19Unfold}. 
In particular, deep unfolding aims at imitating the structure of an iterative optimization algorithm by regarding each iteration as one layer of the neural network, where a number of trainable parameters are introduced to be learned by deep learning techniques \cite{19Unfolding}, as illustrated in Fig. \ref{paradigm}(b). 
For example, the authors of \cite{Shen20LORM} proposed an effective framework for resource management that learns the optimal pruning policy in the conventional branch-and-bound algorithm. In \cite{Hu21DeepUnfold}, a deep unfolding neural network based on the structure of the WMMSE algorithm was developed for precoding in massive MIMO systems. Thanks to the explainable training procedure, preliminary investigations have empirically shown an improved generalization ability of unfolding networks in some cases \cite{Monga21Unroll}, but the computational complexity of such methods is an issue. The iterative nature of the deep unfolding methods makes real-time implementation challenging in large-scale wireless networks. For example, it was shown in \cite{Shen20LORM} that the execution time of the proposed resource management scheme is 2.92 seconds when 10 remote radio heads serve 17 users in a cloud radio access network (Cloud-RAN), which largely exceeds the channel coherence time in practice and results in poor scalability. 
Hence, it can be concluded that scalability and generalization cannot be guaranteed in existing ``learning to communicate” paradigms, which motivates our investigation. 

\subsection{Contributions}
In this paper, a scalable and generalizable system design framework, i.e., neural calibration, is proposed. It can flexibly adapt to different system sizes without retraining the neural network, while enjoying good performance and high computational efficiency. The major contributions are summarized as follows:

\begin{itemize}
	\item We develop an innovative ``learning to communicate” paradigm called \emph{neural calibration} for wireless system design, e.g., channel estimation, signal detection, and resource allocation, which integrates time-efficient model-based algorithms with deep learning. As shown in Fig. \ref{paradigm}(c), two components are involved in the neural calibration framework, i.e., the time-efficient calibration basis and deep learning model. Specifically, the backbone of a low-complexity method is retained while neural networks are adopted to calibrate the input and improve system performance. Different from fully data-driven methods that employ an unstructured neural network, a unique neural network architecture is further developed based on the permutation equivariance property introduced by the wireless network topology. 
	
	\item 	Different from deep unfolding, the domain knowledge we exploit is from a low-complexity algorithm, which enables real-time implementation. By amalgamating the structure of the problem-specific hand-crafted solutions with deep neural networks, the proposed framework achieves satisfactory performance in large-scale wireless systems. These two aspects guarantee the high scalability of the proposed neural calibration scheme. Besides, by exploiting the intrinsic permutation equivariance property of wireless systems, the proposed framework enjoys a good generalization ability when the number of users varies.

	\item 	We apply the proposed neural calibration framework to design downlink beamforming in massive MIMO systems with perfect or implicit channel state information (CSI). Two linear signal processing techniques, i.e., the least-squares (LS) channel estimator and the zero-forcing (ZF) beamformer, will serve as the bases for calibration and are augmented by generalizable neural networks. For the implicit CSI case, the calibrations for channel estimation and beamforming are jointly designed in an end-to-end manner to boost system performance.


	\item Extensive simulations will demonstrate that the performance of the proposed neural calibration-based method matches classic iterative algorithms while enjoying significantly improved scalability and generalization compared with the deep unfolding and the fully data-driven methods. In particular, the proposed neural calibration method achieves $97.6\%$ performance of the WMMSE algorithm with perfect CSI in 0.48 milliseconds when there are 64 antennas and 8 users in massive MIMO systems, while the WMMSE algorithm takes 283.2 milliseconds. Without explicit channel estimates, the neural calibration still obtains $93.1\%$ of the performance upper bound when 128 antennas and 8 users are deployed, while the fully data-driven method only achieves $70.3\%$. 
\end{itemize}

\subsection{Organization} The paper is organized as follows. Section \uppercase\expandafter{\romannumeral2} proposes a general form of the proposed neural calibration framework, where the key properties and the proposed neural network architecture are presented. Section \uppercase\expandafter{\romannumeral3} shows how to implement neural calibration for the downlink beamforming task in massive MIMO systems with perfect CSI and with implicit channel estimation, respectively. Extensive simulations are demonstrated in Section \uppercase\expandafter{\romannumeral4} and conclusions are drawn in Section \uppercase\expandafter{\romannumeral5}.

\subsection{Notations} Scalars, vectors, and matrices are respectively denoted by lower case, boldface lower case, and boldface upper case letters.
$\mathbf{I}$ represents an identity matrix. 
For a matrix $\mathbf{A}$, ${{\bf{A}}^T}$, $\mathbf{A}^*$, ${{\bf{A}}^H}$, and $\|\mathbf{A}\|$ shall denote its transpose, conjugate, conjugate transpose, and Frobenius norm, respectively. ${{\bf{A}}^{-1}}$ denotes the inverse of matrix $\mathbf{A}$.
$\Re\{ \mathbf{A} \}$ and $\Im\{ \mathbf{A} \}$ denote the real and imaginary part of matrix $\mathbf{A}$, respectively. 
Let $\mathbf{\Pi}$ be a permutation matrix. When $\mathbf{\Pi}$ is pre-multiplied or post-multiplied by a matrix $\mathbf{A}$, it results in permuting the rows or columns of $\mathbf{A}$, respectively.
For a vector $\mathbf{a}$, $\|\mathbf{a}\|$ represents its Euclidean norm.
$\mathbb{E}\{ \cdot \}$ denotes the statistical expectation. 
$\operatorname{Tr}(\cdot)$ denotes the trace operation.
$|  \cdot  |$ denotes the absolute value of a complex scalar. 
Finally, ${\mathbb{C}^{m \times n}}\;({\mathbb{R}^{m \times n}})$ denotes the set of ${m \times n}$ complex (real) matrices.

\section{System Model and Problem Formulation} \label{sec:sys_model}

\subsection{System Model}
\begin{figure}[t] 
\centering
\includegraphics[width=0.7\textwidth]{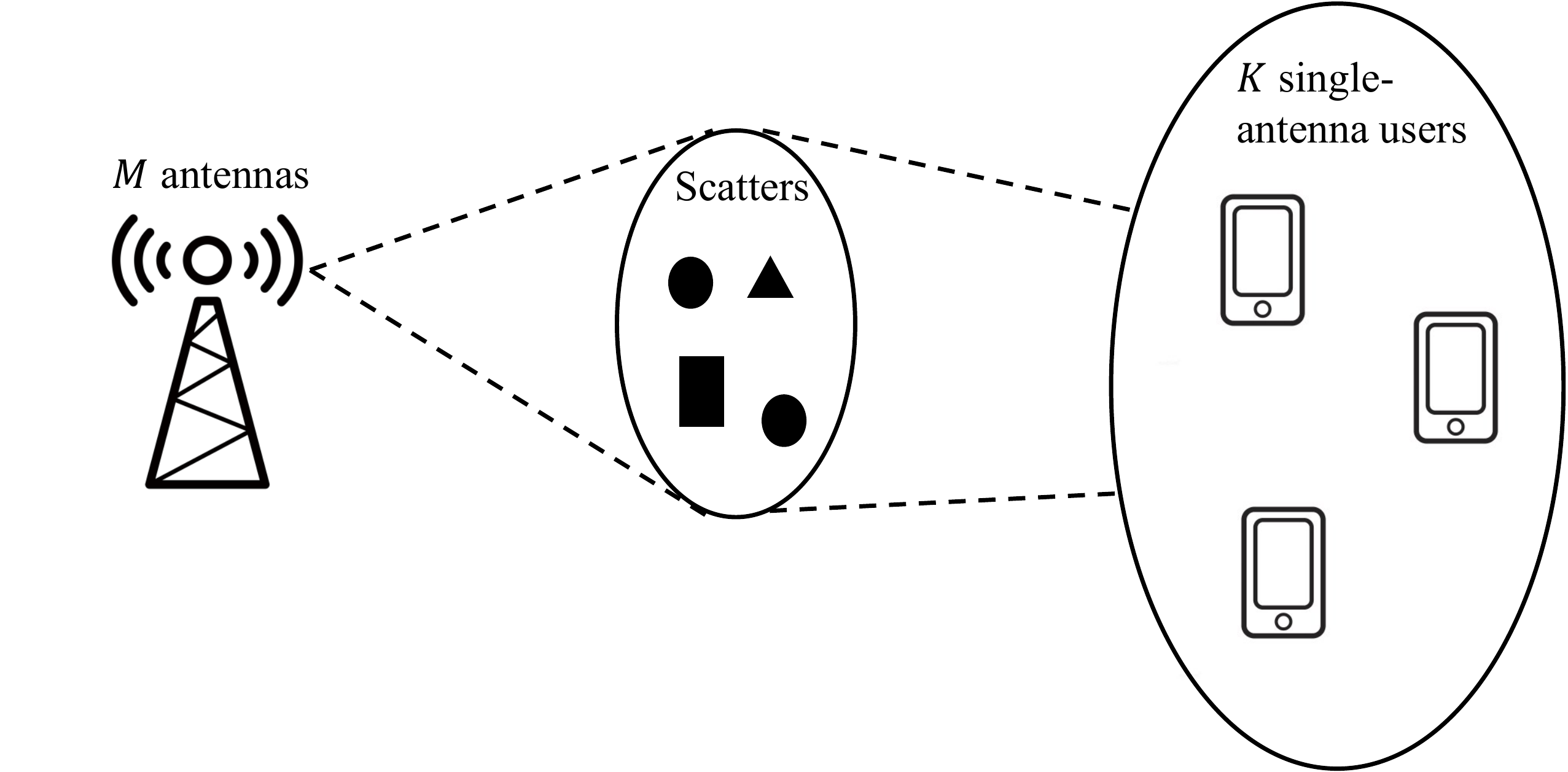} 
\caption{An illustration of the physical communication system.} 
\label{physical_commun} 
\end{figure}

Consider a frequency-division duplexing (FDD) massive MIMO system over a flat block-fading channel\footnote{For ease of illustration, FDD massive MIMO is adopted as the main application example, but the proposed neural calibration framework can be applied to a broad range of system models in wireless communications.} where a base station (BS) with $M$ antennas serves $K$ single-antenna users, as illustrated in Fig. \ref{physical_commun}. We mainly consider downlink data transmission, which will be assisted by uplink feedback signaling. The feedback signal transmitted by $K$ users is denoted as $\mathbf{P} \in\mathbb{C}^{K \times L}$ where $L$ is the length of the feedback symbols. The uplink and downlink channels between user $k$ and the BS are denoted as $\mathbf{h}_{\mathrm{UL},k} \in \mathbb{C}^{M \times 1}$ and $\mathbf{h}_{\mathrm{DL},k} \in \mathbb{C}^{M \times 1}$, respectively. The multi-path channel model \cite{Alrabeiah19FDD} is adopted and the channel between the BS and user $k$ is given by 
\begin{equation}\label{eq:channel_model}
\mathbf{h}_{i,k}=\sum_{\ell=1}^{L_{p}} \alpha_{i, \ell, k} e^{-j 2\pi f_i \tau_{l,k}} \mathbf{a}_{\mathrm{t}}\left(\theta_{\ell, k}\right), \quad i \in \{\mathrm{UL}, \mathrm{DL}\},
\end{equation}
where $L_{p}$ is the number of propagation paths and $f_i$ denotes the carrier frequency. The delay of the $\ell$-th path is represented by $\tau_{l,k}=\frac{d_{l,k}}{c}$, where $c$ is the speed of light and $d_{l,k}$ denotes the distance of the $\ell$-th path. In addition, $\alpha_{i, \ell, k}$ is the complex gain of the $\ell$-th path, $\theta_{\ell,k}$ is the corresponding angle of departure (AoD), and $\mathbf{a}_\mathrm{t}\left(\cdot\right)$ is the transmit array response vector. For a uniform linear array (ULA) with $M$ antenna elements, the transmit array response vector is given by
\begin{equation}
\mathbf{a}_{\mathrm{t}}(\theta)=\left[1, e^{j \frac{2 \pi}{\lambda} d \sin (\theta)}, \ldots, e^{j \frac{2 \pi}{\lambda} d(M-1) \sin (\theta)}\right]^{T},
\end{equation}
where $\lambda$ is the wavelength and $d$ is the antenna spacing. 
For the downlink data transmission phase, let ${\bf{V}}\in \mathbb{C}^{M \times K}$ denote the beamforming matrix at the BS and $\mathbf{s}\in \mathbb{C}^{K\times 1}$ represent the transmit symbol vector. Then, the transmit signal is given by
\begin{equation}
\mathbf{x}=\sum\limits_{k=1}^{K} \mathbf{v}_{k}s_{k}={\bf{V}}\mathbf{s}, 
\end{equation}
where $\mathbf{v}_{k}\in \mathbb{C}^{M \times 1}$ is the $k$-th column of ${\bf{V}}$ and denotes the bearmforming vector for user $k$. The transmit power constraint is therefore given by $\operatorname{Tr}(\mathbf{V} \mathbf{V}^H) \leq P_{\mathrm{DL}}$ with $P_\mathrm{DL}$ denoting the maximum downlink transmit power. The symbol transmitted to user $k$ is denoted by $s_{k}$, where $\mathbb{E}[|s_{k}|^2]=1$ without loss of generality. Then, the received signal at user $k$ is expressed as
\begin{equation}\label{eq_rx_sig}
y_k =  \mathbf{h}_{\mathrm{DL},k}^H \mathbf{v}_{k} s_k + \sum_{j\not=k} \mathbf{h}_{\mathrm{DL},k}^H \mathbf{v}_{j} s_j + n_k,
\end{equation}
where $n_k \sim \mathcal{CN}(0,\sigma^2_0)$ is the additive white Gaussian noise with variance $\sigma^2_0$. 
Therefore, the system sum-rate is given by
\begin{equation}\label{eq_sumrate}
R=\sum\limits_{k=1}^{K} \log_2\left(1 +  \frac{\lvert \mathbf{h}_{\mathrm{DL},k}^H \mathbf{v}_{k} \rvert^2}{ \sum_{j\not=k} \lvert \mathbf{h}_{\mathrm{DL},k}^H \mathbf{v}_{j} \rvert^2+\sigma^2_0 } \right).
\end{equation}

\subsection{Problem Formulation}
To fully exploit the potential of the considered massive MIMO system, different design tasks can be carried out for performance optimization. 
Optimization problems can be generally formulated as
\begin{equation} \label{general}
\min _{\mathbf{X}} \quad f(\mathbf{X} ; \mathbf{Z}) \quad \text { s.t. } \mathbf{X} \in \mathcal{X},
\end{equation}
where $\mathbf{X}$ is the optimization variable, $f(\cdot) : \mathbb{C}^{m \times n} \rightarrow \mathbb{R}$ is the objective function, $\mathcal{X}$ is the feasible set, and $\mathbf{Z} \in \mathbb{C}^{p \times q}$ denotes the given parameter of the problem. Several typical optimization problems in wireless systems and their corresponding forms of Problem \eqref{general} are shown in Table \ref{problem_example}. 
In the following, we illustrate two instances of Problem \eqref{general} in detail, i.e., the uplink channel estimation and downlink beamforming design. 

\begin{table*}[t]	
\selectfont  
\centering
\newcommand{\tabincell}[2]{\begin{tabular}{@{}#1@{}}#2\end{tabular}}
\caption{Typical Examples of the General Formulation \eqref{general}.} 
\begin{tabular}{|c|c|c|c|}
\hline
\textbf{Problem} & \textbf{Objective} & \textbf{Set constraint $\mathcal{X}$} & \textbf{System Parameter $\mathbf{Z}$} \\ \hline
Channel estimation & \tabincell{c}{Minimize expected \\ estimation error} & $-$ & Pilot symbols and received pilots  \\ \hline
Signal detection & \tabincell{c}{Maximize likelihood function} & Constellation set constraint
& \tabincell{c}{Received signals and \\ estimated channels  } \\ \hline
Beamforming & \tabincell{c}{Maximize system \\ spectral efficiency} & 
Total transmit power constraint
& Downlink CSI and noise power \\ \hline
Power allocation & \tabincell{c}{Maximize system capacity} & 
Total transmit power constraint
& CSI and noise power  \\ \hline
\end{tabular}
\label{problem_example}
\end{table*}

In order to estimate the uplink channel matrix $\mathbf{H}_{\mathrm{UL}} = [\mathbf{h}_{\mathrm{UL},1}, \cdots, \mathbf{h}_{\mathrm{UL},K}] \in \mathbb{C}^{M \times K}$, $L$ training pilots are transmitted by $K$ users. In this case, $\mathbf{P} \in \mathbb{C}^{K \times L}$ denotes uplink pilots. The received pilot matrix is given by
\begin{equation}
\mathbf{Y}_{\mathrm{p}} = \mathbf{H}_{\mathrm{UL}} \mathbf{P} + \mathbf{N},
\end{equation}
where $\mathbf{N} \in \mathbb{C}^{M \times L}$ is the additive white Gaussian noise matrix. The task of the uplink channel estimation is to recover the channel matrix $\mathbf{H}_{\mathrm{UL}}$ based on the knowledge of $\mathbf{Y}$ and $\mathbf{P}$, which is formulated as
\begin{equation}\label{CE}
\begin{aligned}
\min_{\mathcal{P}(\cdot)} ~& \mathbb{E}\left\{\| \mathbf{H}_{\mathrm{UL}} - \mathcal{P}(\mathbf{Y}_{\mathrm{p}}; \mathbf{P})\|^{2}\right\},
\end{aligned}
\end{equation}
where $\mathcal{P}$ denotes the channel estimation scheme that finds $\hat{\mathbf{H}}_{\mathrm{UL}}=\mathcal{P}(\mathbf{Y}_{\mathrm{p}}; \mathbf{P})$. 

For the downlink beamforming design, we assume that perfect downlink CSI is available. To maximize the system sum-rate, the downlink beamforming problem is formulated as
\begin{equation}
\begin{aligned}
\label{BF_perfectCSI}
\max_{\mathbf{V}} ~ & \sum_{k=1}^{K} \log_2\left(1 +  \frac{\lvert \mathbf{h}_{\mathrm{DL},k}^H \mathbf{v}_k \rvert^2}{ \sum_{j\not=k} \lvert \mathbf{h}_{\mathrm{DL},k}^H \mathbf{v}_j \rvert^2+\sigma^2_0 } \right)\\ 
\text{s.t.} ~ & \operatorname{Tr}(\mathbf{V} \mathbf{V}^H) \leq P_{\mathrm{DL}}.
\end{aligned}
\end{equation}

\subsection{Existing Approaches}
\subsubsection{Fully Data-Driven Methods} In order to solve Problem \eqref{general}, fully data-driven methods \cite{sun2018learning, liang2018towards, lee2018deep} 
have been developed with the following general expression
\begin{equation}
\hat{\mathbf{X}}=\mathcal{Q}\left(\mathbf{Z}; \bm{\Phi} \right).
\end{equation}
In principle, the fully data-driven approach aims at learning the complicated mapping $\mathcal{Q}$ from the given parameter $\mathbf{Z}$ to the optimized solution $\hat{\mathbf{X}}$ via a neural network parameterized by $\mathbf{\Phi}$. Note that the computationally-demanding training stage is generally done offline, and the online inference stage only requires several layers of simple operations such as matrix multiplications. Therefore, these fully data-driven methods support real-time implementation \cite{sun2018learning}. 
However, by adopting black-box neural networks, the underlying high-dimensional mapping $\mathcal{Q}$ is difficult to learn in large-scale wireless systems, which leads to poor scalability and generalization abilities.

\subsubsection{Deep Unfolding} To exploit the domain knowledge inherited in conventional model-based algorithms, deep unfolding methods were proposed to imitate the iterative procedure of the derived algorithms \cite{19Unfolding}. 
For a general iterative algorithm, the operation in each iteration can be characterized by
\begin{equation}\label{iter}
\mathbf{X}^{t} = F_t (\mathbf{X}^{t-1}; \mathbf{Z}), \quad t \in \{1, 2, \ldots, T\},
\end{equation}
where $T$ denotes the total number of iterations and the function $F_t$ maps the variable $\mathbf{X}^{t-1}$ to $\mathbf{X}^{t}$ in the $t$-th iteration based on the parameter $\mathbf{Z}$. To improve the performance for a fixed number of iterations, the function $F_t$ can be replaced as one layer of a neural network. As such, the iteration expressed in \eqref{iter} is transformed into \cite{Hu21DeepUnfold}
\begin{equation}
\mathbf{X}^{n} = \mathcal{F}_n (\mathbf{X}^{n-1}; {\bm{\theta}}^{n}, \mathbf{Z}), \quad n \in \{1, 2, \ldots, N\},
\end{equation}
where $n$ denotes the layer index in the neural network, $N$ is the total number of layers, ${\bm{\theta}}^{n}$ represents the trainable parameters in the $n$-th layer, $\mathcal{F}_n$ is the non-linear mapping in the $n$-th layer of the neural network, and $\mathbf{X}^{n-1}$ and $\mathbf{X}^{n}$ represent the input and output of the $n$-th layer, respectively. Although the deep unfolding algorithm is self-explainable, it imitates the structure of the conventional iterative algorithm, making it challenging to be executed in real-time in large-scale wireless networks \cite{Hu21DeepUnfold}. 

\section{Proposed Neural Calibration Framework} \label{sec:framework}
In this section, we propose a neural calibration framework that integrates time-efficient model-based techniques with deep learning to solve Problem \eqref{general}. The function derived from the time-efficient algorithm serves as the basis for calibration. In addition, the fundamental topological structure of wireless systems is exploited to identify a permutation equivariance property, based on which a scalable and generalizable neural network architecture is developed.

\subsection{General Framework}
We propose a new paradigm for ``learning to communicate”, i.e., the neural calibration framework, where we combine the advantages of the classic model-based algorithms and data-driven methods. In particular, the backbone of the model-based method is kept in place as the basis of calibration, and neural networks are adopted to calibrate the inputs to the classic modules. Mathmatically, the neural calibration approach is given by
\begin{equation} \label{NC}
\hat{\mathbf{X}}=g_{\mathrm{TE}}(\mathcal{L}\left(\mathbf{Z}; \mathbf{\Psi} \right)),
\end{equation}
where $g_{\mathrm{TE}}$ denotes the function derived from the corresponding model-based algorithm, $\mathcal{L}$ represents the deep learning model used to learn the calibration scheme, and $\mathbf{\Psi}$ is the trainable parameter in the learning component. According to \eqref{NC}, the system parameter $\mathbf{Z}$ is inputted to the learning model $\mathcal{L}$ parameterized by $\mathbf{\Psi}$. Then, the output of the learning model is fed as the input of the function $g_\mathrm{TE}$ to generate the final solution $\hat{\mathbf{X}}$. In the following, we shall introduce the two components in the proposed framework in detail, i.e., the calibration basis algorithm, $g_\mathrm{TE}$, and the neural network architecture, $\mathcal{L}$.

\subsection{Neural Calibration: Time-Efficient Basis}
Recall that, for the deep unfolding approach, the object to unfold is an iterative model-based algorithm, which leads to the failure in achieving high scalability. Hence, to facilitate the scalability of the proposed neural calibration framework, we adopt simple but powerful time-efficient algorithms as the basis of calibration, whose input is further tunned by the neural network. For example, it has been shown that time-efficient approaches such as the matched filter (MF) detector, ZF beamformer, and linear minimum mean-squared error (LMMSE) estimator are asymptotically optimal when the number of antennas at the BS is large enough compared to the number of users \cite{Marzetta10MIMO, Rusek13MIMO}. 
Typical problems and the corresponding time-efficient methods are shown in Table \ref{time-efficient}. 

\begin{table*}[t]	
\selectfont  
\centering
\newcommand{\tabincell}[2]{\begin{tabular}{@{}#1@{}}#2\end{tabular}}
\caption{Typical Problems and Corresponding Time-efficient Bases.} 
\begin{tabular}{|c|c|}
\hline
\textbf{Problem} & \textbf{Time-efficient methods} \\ \hline
Channel estimation & LS estimator, scaled LS estimator, and LMMSE estimator \cite{Biguesh06CE}  \\ \hline
Signal detection & MF detector, ZF detector, and LMMSE detector\cite{Yang15detection}   \\ \hline
Beamforming design & \tabincell{c}{Maximum ratio transmission (MRT) beamformer, ZF beamformer, \\ and regularized ZF beamformer \cite{Lu14MIMO}}   \\ \hline
Power allocation & Fast water filling method \cite{Ling12FWF}  \\ \hline
\end{tabular}
\label{time-efficient}
\end{table*}

Although the conventional time-efficient method $ g_{\mathrm{TE}} (\mathbf{Z})$ benefits from high computational efficiency, it suffers from performance degradation in the low SNR regime or when the number of users has nearly the same order of magnitude as that of antennas in a massive MIMO system. This indicates that the inputs of these time-efficient functions are not optimal. Therefore, instead of directly feeding the system parameter $\mathbf{Z}$ into the function $g_{\mathrm{TE}}$, we adopt a deep neural network to calibrate the input of the conventional time-efficient function.

\subsection{Neural Calibration: Deep Learning Component}
In existing fully data-driven approaches, the input and output dimensions of the learning model $\mathcal{Q}$ are typically proportional to the number of antennas and the number of users, which are large in modern communication systems. Nevertheless, simply employing giant and unstructured neural networks does not incorporate the uniqueness of wireless problems and thus, suffers from poor scalability and generalization. Therefore, a more delicate design of the learning component in neural calibration is desired. 
In this subsection, we show two permutation equivariance properties in the uplink and downlink transmissions in wireless systems, i.e., the antenna-wise permutation equivariance and the user-wise permutation equivariance, respectively, based on which the generalizable neural network architecture of the learning component in the proposed neural calibration framework is designed.


\subsubsection{Uplink Antenna-Wise Permutation Equivariance}
In uplink signal processing tasks, one crucial feature is the antenna-wise permutation equivariance. To make our discussion concrete, we take the uplink channel estimation as an example. Recall that the uplink channel estimation problem is formulated as in \eqref{CE}.
Let $\mathbf{\Pi}$ denote the permutation matrix and $\mathbf{H}_{\mathrm{UL}}^{'}$ represent the permutated channel matrix, where the rows are permuted according to $\mathbf{\Pi}$, i.e., $\mathbf{H}_{\mathrm{UL}}^{'} = \mathbf{\Pi}^T \mathbf{H}_{\mathrm{UL}}$. Similarly, ${\mathbf{Y}}_\mathrm{p}^{'}$ is the permuted pilot matrix according to the same permutation $\mathbf{\Pi}$, i.e., ${\mathbf{Y}}_\mathrm{p}^{'} = \mathbf{\Pi}^T {\mathbf{Y}_\mathrm{p}}$. We now prove the antenna index permutation equivariance of Problem \eqref{CE} as stated below, which implies the ordering of the antenna elements has no impact on the network performance. 

\begin{thm}\label{UplinkPE}
(Antenna-Wise Permutation Equivariance) The channel estimator has the following property 
\begin{equation}\label{AntennaWisePE}
\mathcal{P}(\mathbf{\Pi}^T {\mathbf{Y}_\mathrm{p}}) = \mathbf{\Pi}^T \mathcal{P}({\mathbf{Y}_\mathrm{p}}),
\end{equation}
for any permutation matrix $\mathbf{\Pi}$.
\end{thm}
\begin{IEEEproof}
Please refer to Appendix \ref{proof:UplinkPE}.
\end{IEEEproof}

It is implied in \eqref{AntennaWisePE} that, irrespective of whether we first permute the received pilot matrix and then estimate the channel or perform the two steps in the reverse order, we eventually get the same results. This implies that all the rows of $\mathbf{Y}_\mathrm{p}$ are homogeneous and how the ordering of these rows is presented to the estimator should have no impact on the estimation error.

For the learning component design in the calibration framework, inspired by Theorem \ref{UplinkPE}, we develop $M$ duplicate MLPs that share the common trainable parameters for $M$ antenna elements to approximate the mapping $\mathcal{L}$, as shown in Fig. \ref{FrameworkStructure}(a).
The input and output of each MLP are the $i$-th row of the received pilot matrix $\mathbf{Y}_\mathrm{p}$ and the calibrated results $\widetilde{\mathbf{Y}}_\mathrm{p}$, respectively, whose dimensions are both reduced from $2LM$ to $2L$, which is independent of the number of received antennas and therefore improves the scalability and generalization of the proposed neural calibration framework. 

With the calibration basis in Table \ref{time-efficient} and the generalizable neural network architecture in Fig. \ref{FrameworkStructure}(a) at hand, the channel estimation problem \eqref{CE} is reformulated as 
\begin{equation} \label{NC_CE}
\min_{\bm{\Psi}} ~ \mathbb{E}\left\{\| \mathbf{H}_{\mathrm{UL}} - g_{\mathrm{CE}}(\mathcal{L}(\mathbf{Y}_\mathrm{p}; \bm{\Psi}); \mathbf{P})\|^{2}\right\},
\end{equation}
where $g_{\mathrm{CE}}$ denotes the function derived from the time-efficient channel estimator, e.g., $g_{\mathrm{CE}}(\mathbf{Y}_\mathrm{p}; \mathbf{P}) \\ = \mathbf{Y}_\mathrm{p} \mathbf{P}^H {(\mathbf{P} \mathbf{P}^H)}^{-1}$ according to the LS estimator. Owing to the similarities between the channel estimation and the MIMO detection problem \cite{Yang15detection}, the antenna-wise permutation equivariance can be naturally transplanted to the design of an effective data detector.

\begin{figure}[t] 
\centering
\includegraphics[width=0.85\textwidth]{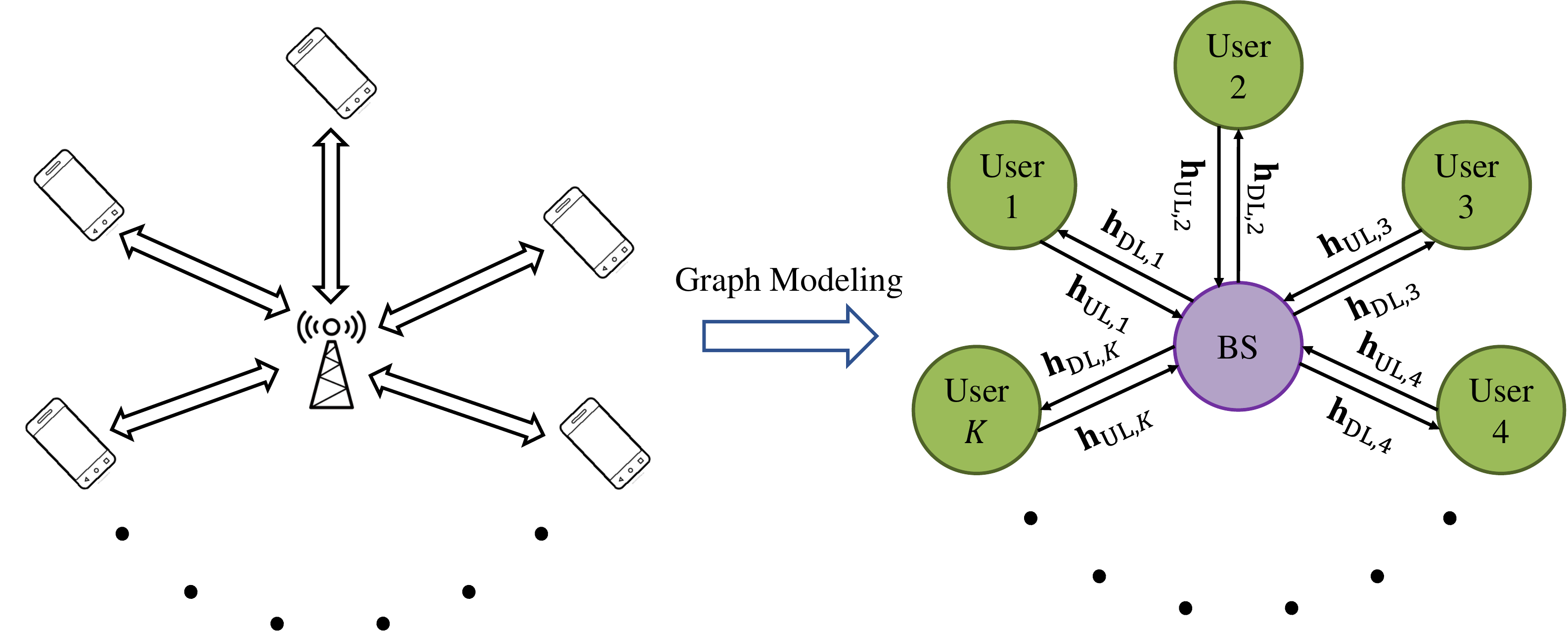} 
\caption{An illustration of graph modeling of a single cell communication system.} 
\label{graph_modeling} 
\end{figure}
\subsubsection{Downlink User-Wise Permutation Equivariance}
Similarly, for the downlink signal processing tasks, we can develop a scalable and generalizable neural network architecture by identifying the user-wise permutation equivariance property, i.e., the ordering of the users does not count when optimizing the wireless systems. To this end, we resort to graph theory, which has found abundant successful applications such as radio resource management \cite{Shen21Graph} and MIMO detection \cite{Scotti20GNN}. 
We take the downlink resource allocation as an example. 
The wireless network can be modeled as a directed graph with node and edge features \cite{Shen21Graph}. In particular, each transmitter or receiver is modeled as one node in the graph, and an edge is drawn from node $i$ to node $j$ if there is a transmission link from transmitter $i$ to receiver $j$. An illustration of such a graph modeling is shown in Fig. \ref{graph_modeling}. The node features include the properties of the node, e.g., the weights of the users in the weighted sum-rate maximization problem. Let $\mathbf{N} \in \mathbb{C}^{n \times d_1}$ denote the node feature matrix, where $n$ is the total number of nodes and $d_1$ is the dimension of the node feature. The edge features consist of the properties of the corresponding channels, e.g., instantaneous CSI. Let $\mathbf{A} \in \mathbb{C}^{n \times n \times d_2}$ denote the edge feature tensor, where $d_2$ is the dimension of the edge feature. We assign each node an optimization variable $\gamma_i \in \mathbb{C}^{c \times 1}$ and let $\mathbf{\Gamma} = [\gamma_1, \cdots, \gamma_n]^{T} \in \mathbb{C}^{n \times c}$.
It is shown in \cite{Shen21Graph} that resource allocation problems can be written as an optimization over the wireless channel graph, which is given by
\begin{equation} \label{user-wisePE}
\min _{\mathbf{\Gamma}} \quad g(\mathbf{\Gamma} ; \mathbf{N}, \mathbf{A}) \quad \text { s.t. } \mathbf{\Gamma} \in \mathcal{Y},
\end{equation}
where $g$ denotes the objective function and $\mathcal{Y}$ denotes the constraint set, e.g., transmit power constraint. Let $\mathcal{R}$ be the resource allocation scheme that maps the system parameters to the solutions, i.e., $\mathbf{\Gamma} = \mathcal{R}(\mathbf{N}, \mathbf{A})$. According to \cite[Prop. 4]{Shen21Graph}, the user-wise permutation equivariance property holds for Problem \eqref{user-wisePE}, which is stated below.
\begin{prop} \label{user-wisePerm}  (User-Wise Permutation Equivariance)
The resource allocation scheme $\mathcal{R}$ has the following property
\begin{equation}
\mathcal{R}(\mathbf{\Pi}^T {\mathbf{N}}, \mathbf{\Pi}^T {\mathbf{A}} \mathbf{\Pi}) = \mathbf{\Pi}^T \mathcal{R}({\mathbf{N}}, \mathbf{A}),
\end{equation}
for any permutation matrix $\mathbf{\Pi}$.
\end{prop}
\begin{IEEEproof}
Please refer to \cite[Prop. 4]{Shen21Graph}.
\end{IEEEproof}

\begin{figure}[t] 
\centering
\subfigure[Uplink antenna-wise permutation equivariance.]{
\begin{minipage}[t]{0.5\linewidth}
\centering
\includegraphics[width=2.8in]{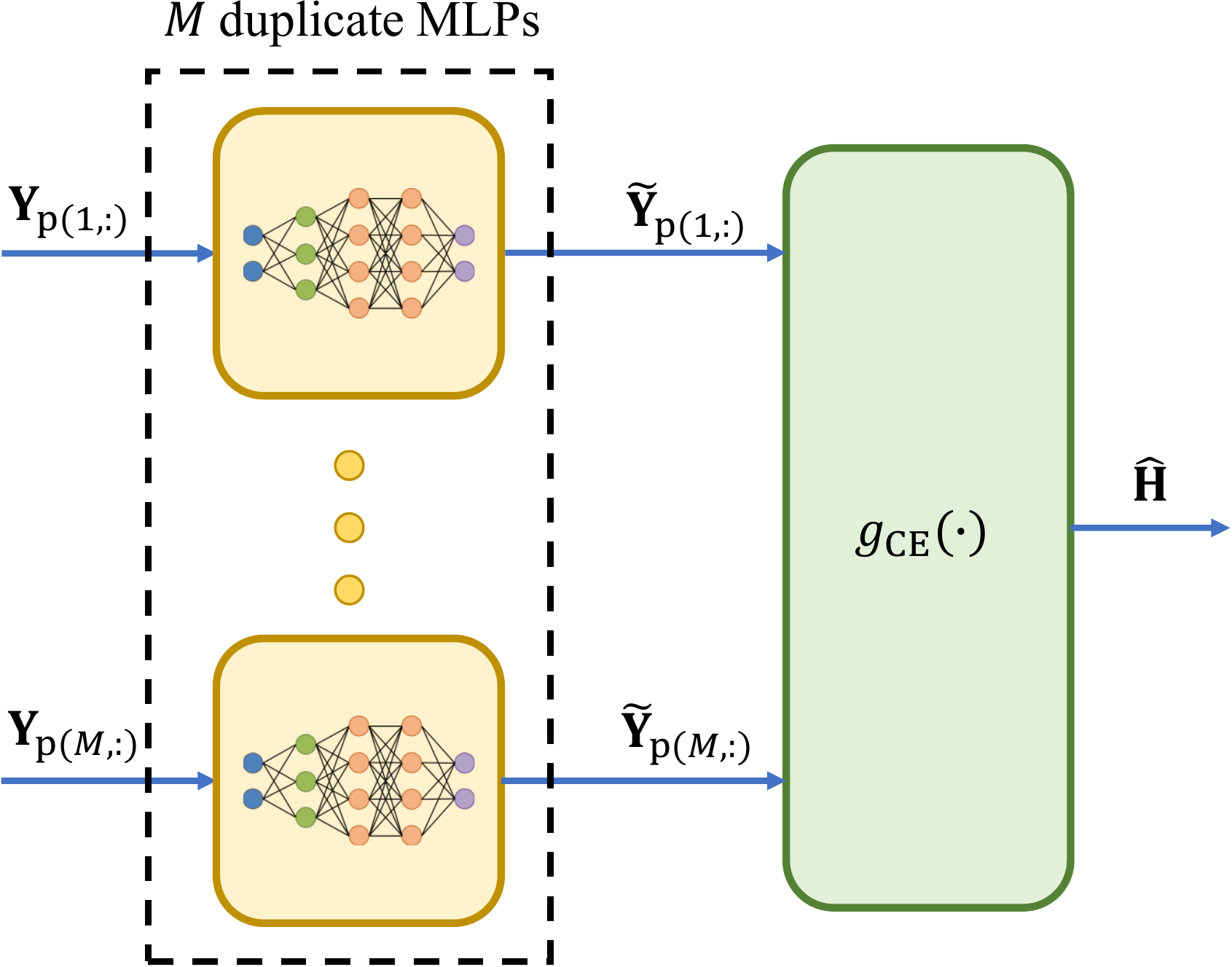}
\end{minipage}%
}%
\subfigure[Downlink user-wise permutation equivariance.]{
\begin{minipage}[t]{0.5\linewidth}
\centering
\includegraphics[width=2.8in]{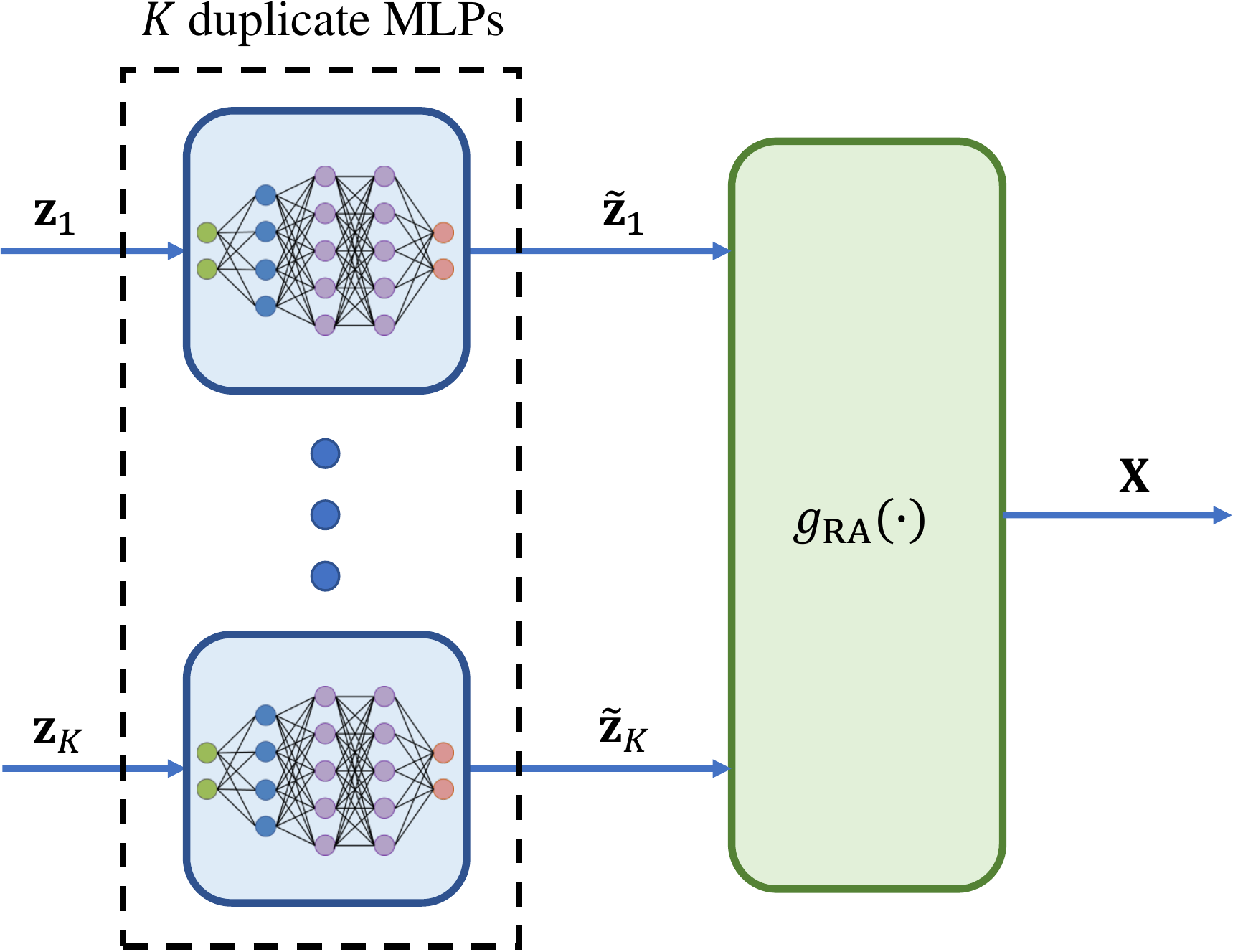}
\end{minipage}%
}%
\centering
\caption{Generalizable neural network architecture design in neural calibration.}
\label{FrameworkStructure}
\end{figure}

Proposition \ref{user-wisePerm} indicates that the ordering of the users presented to the solver $\mathcal{R}$ should have no impact on system performance. This irrelevance of the user ordering inspires us to develop $K$ duplicate MLPs that share the same trainable parameters for $K$ users to learn the calibration scheme $\mathcal{L}$ in \eqref{NC}. The corresponding structure is shown in Fig. \ref{FrameworkStructure}(b), where $\mathbf{z}_k$ denotes the system parameter related to the $k$-th user, $\widetilde{\mathbf{z}}_k$ denotes the calibrated result, and $g_{\mathrm{RA}}$ denotes the function derived from the time-efficient resource allocation solution. The input and output dimensions of each MLP are independent of the number of users, which enables a good generalization ability when the number of users changes.

\section{Neural Calibration for Beamforming Design} \label{sec:BF}
To demonstrate the effectiveness of the proposed framework, in this section we present two applications in wireless communication systems. Specifically, the downlink beamforming designs in massive MIMO systems with perfect CSI and with implicit channel estimation are considered.


\subsection{Beamforming Design with Perfect Downlink CSI}
Recall that given a perfect downlink CSI, the beamforming problem is formulated as in \eqref{BF_perfectCSI}.
Problem \eqref{BF_perfectCSI} is nonconvex with respect to the beamforming matrix $\mathbf{V}$, making the direct optimization of \eqref{BF_perfectCSI} challenging. 
In the following, we introduce how to implement the neural calibration-based method to solve Problem \eqref{BF_perfectCSI}. 
In particular, we adopt the ZF beamformer as the calibration basis, i.e., $h_{\mathrm{ZF}}(\mathbf{H}_{\mathrm{DL}}) = \gamma_{\mathrm{ZF}} \mathbf{H}_{\mathrm{DL}}^{H}(\mathbf{H}_{\mathrm{DL}} \mathbf{H}_{\mathrm{DL}}^{H})^{-1}$ where $\gamma_{\mathrm{ZF}}$ is the scaling constant to satisfy the power constraint and $\mathbf{H}_{\mathrm{DL}} = [\mathbf{h}_{\mathrm{DL},1}, \ldots, \mathbf{h}_{\mathrm{DL},K}]^H \in \mathbb{C}^{K \times M}$. According to \eqref{NC}, the input of the ZF beamformer is calibrated by neural networks. 
The neural calibration-based downlink beamforming design is then reformulated as
\begin{equation}
\begin{aligned}
\label{BF_perfectCSI_NC}
\max_{\mathcal{F}(\cdot)} ~ & \sum_{k=1}^{K} \log_2\left(1 +  \frac{\lvert \mathbf{h}_{\mathrm{DL},k}^H \mathbf{v}_k \rvert^2}{ \sum_{j\not=k} \lvert \mathbf{h}_{\mathrm{DL},k}^H \mathbf{v}_j \rvert^2+\sigma^2_0 } \right)\\ 
\text{s.t.} ~ & \mathbf{V} = h_{\mathrm{ZF}}(\mathcal{F}(\mathbf{H}_{\mathrm{DL}})),
\end{aligned}
\end{equation}
where $\mathcal{F}$ denotes the calibration scheme of the ZF beamformer. We first prove that the calibration mapping $\mathcal{F}$ exists.

\begin{thm}\label{Potential}
(Potential to Calibrate) Define 
\begin{align*}
    &\mathbf{V}_{\mathrm{ZF}} = h_{\mathrm{ZF}}(\mathbf{X}) = \gamma_{\mathrm{ZF}} \mathbf{X}^{H}\left(\mathbf{X} \mathbf{X}^{H}\right)^{-1} \text{and} \\
    &R(\mathbf{V}_{\mathrm{ZF}}) = \sum_{k=1}^{K} \log_2\left(1 +  \frac{\lvert \mathbf{h}_{\mathrm{DL},k}^H \mathbf{v}_{\mathrm{ZF},k} \rvert^2}{ \sum_{j\not=k} \lvert \mathbf{h}_{\mathrm{DL},k}^H \mathbf{v}_{\mathrm{ZF},j} \rvert^2+\sigma_0^2 } \right).
\end{align*}
Then, when $\sigma_0^2 \neq 0$, $\forall \mathbf{H}_{\mathrm{DL}}$ and $\forall \epsilon > 0$, there exists an $\widetilde{\mathbf{H}}_{\mathrm{DL}}$ with $\|\mathbf{H}_{\mathrm{DL}} - \widetilde{\mathbf{H}}_{\mathrm{DL}}\| \leq \epsilon$ such that $R\left(h_{\mathrm{ZF}}(\mathbf{H}_{\mathrm{DL}})\right) < R(h_{\mathrm{ZF}}(\widetilde{\mathbf{H}}_{\mathrm{DL}}))$.
\end{thm}
\begin{IEEEproof}
Please refer to Appendix \ref{proof:Potential}.
\end{IEEEproof}

Theorem \ref{Potential} indicates that, for every channel realization $\mathbf{H}_{\mathrm{DL}}$, there always exists a beamformer $h_{\mathrm{ZF}}(\widetilde{\mathbf{H}}_{\mathrm{DL}})$ other than the conventional ZF beamformer $h_{\mathrm{ZF}}(\mathbf{H}_{\mathrm{DL}})$, which achieves a higher system sum-rate.
Note that even if it exists, it is typically unknown and hard to analytically characterize. 
In this case, the powerful learning capabilities of MLPs can be leveraged to approximate the complicated mapping $\widetilde{\mathbf{H}}_{\mathrm{DL}} = \mathcal{F}(\mathbf{H}_{\mathrm{DL}})$. 

As discussed in Section \uppercase\expandafter{\romannumeral3}, for this downlink resource allocation problem, $K$ duplicate MLPs with the same trainable parameters are adopted for $K$ different users to approximate $\mathcal{F}$. 
The structure shown in Fig. \ref{FrameworkStructure}(b) is utilized, where $\mathbf{z}_k$ is implemented by $\mathbf{h}_{\mathrm{DL},k}$, $g_{\mathrm{RA}}$ is implemented by $h_{\mathrm{ZF}}$, and the output of the whole module is the optimized beamforming matrix $\mathbf{V}$.
Therefore, $\mathcal{F}(\mathbf{H}_{\mathrm{DL}}) = [\phi (\mathbf{h}_{\mathrm{DL},1}), \ldots, \phi(\mathbf{h}_{\mathrm{DL},K})]^H$ where $\phi$ denotes the parameterized function of the MLP.
Let $\mathbf{\Pi}$ denote the permutation matrix. The permutation of the user index implies that the rows of $\mathbf{H}_{\mathrm{DL}}$ is permutated by $\mathbf{\Pi}$, i.e., $\mathbf{\Pi}^{T} \mathbf{H}_{\mathrm{DL}}$, and the columns of $\mathbf{V}$ are permutated by the same permutation matrix, i.e., $\mathbf{V} \mathbf{\Pi}$.
We then show that the proposed neural calibration-based beamformer satisfies the user-wise permutation equivariance property in Proposition \ref{perm}.

\begin{prop} \label{perm}
The proposed neural calibration-based beamformer $h_{\mathrm{ZF}}(\mathcal{F}(\mathbf{H}_{\mathrm{DL}}))$ satisfies the user-wise permutation equivariance property
\begin{equation}
h_{\mathrm{ZF}}(\mathcal{F}(\mathbf{\Pi}^{T} \mathbf{H}_{\mathrm{DL}})) = h_{\mathrm{ZF}}(\mathcal{F}(\mathbf{H}_{\mathrm{DL}})) \mathbf{\Pi}.
\end{equation}
\end{prop}
\begin{IEEEproof}
Please refer to Appendix \ref{proof:perm}.
\end{IEEEproof}

Proposition \ref{perm} indicates that, unlike the fully data-driven method that learns the user-wise permutation equivariance property from a large amount of data, the proposed neural calibration-based method explicitly exploits the equivariance property.
For each channel realization, the structure of duplicate MLPs makes the sample being reused for $K$ times. Therefore, the neural calibration-based method requires fewer training samples compared with the fully data-driven methods. It is shown in \cite{shen2021ai} that a high sample efficiency results in a small optimality gap, implying that the proposed method enjoys a higher scalability. Besides, since the input and output dimensions of each MLP are irrelevant to the number of users, the proposed neural calibration-based method can be applied without re-training when the user number varies and thus, achieves higher generalization. In practice, the downlink CSI is unknow and needs to be estimated. In the next subsection, we consider a more challenging application where the channel estimation module and the beamforming module are jointly designed to boost the system performance.

\subsection{Beamforming Design with Implicit Channel Estimation}
In practice, downlink CSI acquisition is a challenging task, especially in FDD massive MIMO systems due to the weak reciprocity between the uplink and downlink channels. The commonly-adopted solution lets users estimate the downlink channel first, and then send the information back to the BS for beamforming \cite{KNLau14}, which causes prohibitive delay and overhead. Since the uplink and downlink channels share the same physical propagation environment, it was shown in \cite{Alrabeiah19FDD} that an intricate mapping from the uplink to the downlink exists and can be approximated by a well-trained MLP. 

Conventionally, channel estimation and beamforming are treated as two stand-alone components, which makes it difficult to achieve a global system optimality \cite{Ye20E2E}. Recently, deep learning-based approaches for beamforming design with implicit channel estimation were proposed in \cite{Sohrabi21FDD, Jiang20Implicit}, where the beamformers at the BS are directly optimized according to the received pilots. 
In these methods, the BS extracts the useful information from the received pilots $\mathbf{Y}_{\mathrm{p}}$ and then directly design the downlink beamforming $\mathbf{V}$, i.e.,
\begin{equation} \label{functionG}
\mathbf{V} = \mathcal{G}\left( \mathbf{Y}_{\mathrm{p}} \right) = \mathcal{G}\left( \mathbf{H}_{\mathrm{UL}} \mathbf{P} + \mathbf{N} \right),
\end{equation}
where $\mathcal{G}(\cdot): \mathbb{C}^{M\times L} \rightarrow \mathbb{C}^{M\times K} $ represents the mapping from the received pilots to the downlink beamformer. Note that this method bypasses the explicit channel estimation stage and directly optimizes the beamforming matrix. Thus, the sum-rate maximization problem with implicit channel estimation is formulated as
\begin{equation}
\begin{aligned}
\label{BF_implicit}
\max_{\mathcal{G}(\cdot)} ~ & \mathbb{E} \left \{ \sum_{k=1}^{K} \log_2\left(1 +  \frac{\lvert \mathbf{h}_{\mathrm{DL},k}^H \mathbf{v}_k \rvert^2}{ \sum_{j\not=k} \lvert \mathbf{h}_{\mathrm{DL},k}^H \mathbf{v}_j \rvert^2+\sigma_0^2 } \right) \right \}\\ 
\text{s.t.} ~ & \mathbf{V} = \mathcal{G}\left( \mathbf{Y}_{\mathrm{p}} \right) = \mathcal{G}\left( \mathbf{H}_{\mathrm{UL}} \mathbf{P} + \mathbf{N} \right),\\
& \operatorname{Tr}(\mathbf{V} \mathbf{V}^H) \leq P_{\mathrm{DL}},
\end{aligned}
\end{equation}
where the expectation is over the random channel realizations and the noise in the uplink pilot transmission phase. Note that there is no model-based algorithm to solve \eqref{BF_implicit} since the mapping $\mathcal{G}$ is complicated and hard to characterize. 
However, utilizing a black-box neural network trained with a large amount of data to approximate the mapping $\mathcal{G}$ suffers from poor scalability when the wireless network size increases and is deficient in generalization ability.
In contrast, we will show how to implement neural calibration to solve \eqref{BF_implicit} with high scalability and generalization. 

\begin{figure}[t] 
\centering
\includegraphics[width=0.8\textwidth]{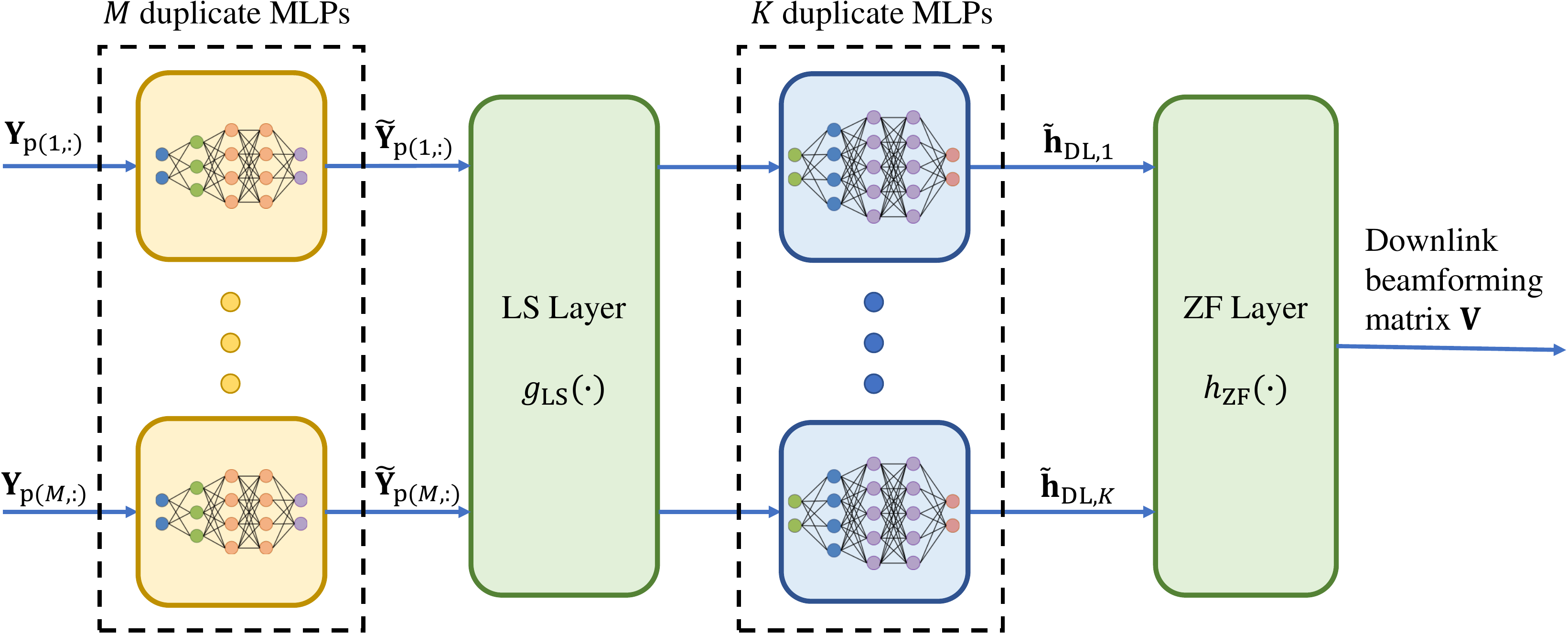} 
\caption{Proposed neural calibration-based end-to-end design architecture for Problem \eqref{BF_implicit}.} 
\label{section3} 
\end{figure}
The overall block diagram of the proposed neural calibration-based end-to-end design in the training stage is shown in Fig. \ref{section3}. In particular, we adopt the classic LS estimator and ZF beamformer, i.e., $g_{\mathrm{LS}}$ and $h_{\mathrm{ZF}}$, as the basis for calibration. Thanks to the universal approximation property, neural networks are adopted to calibrate the inputs of the linear functions to improve the performance of the end-to-end design. Then, the neural calibration-based downlink beamforming optimization with implicit downlink channel estimation is reformulated as
\begin{equation} \label{CM+CaliZF}
\begin{aligned}
\max_{\mathcal{F} \circ \mathcal{H}(\cdot), \mathcal{B}(\cdot)} ~& \sum_{k=1}^{K} \log_2\left(1 +  \frac{\lvert \mathbf{h}_{\mathrm{DL},k}^H \mathbf{v}_k \rvert^2}{ \sum_{j\not=k} \lvert \mathbf{h}_{\mathrm{DL},k}^H \mathbf{v}_j \rvert^2+\sigma_0^2 } \right)\\ 
\text{s.t.}  ~ & \mathbf{V} = h_{\mathrm{ZF}}( \mathcal{F} \circ \mathcal{H}(f_{\mathrm{LS}}(\mathcal{B}(\mathbf{P})) ) ), \\
&\|\widetilde{\mathbf{x}}_k\|_2^2\leq P_{\mathrm{UL}} L, ~~\forall k,
\end{aligned}
\end{equation}
where $\mathcal{B}(\cdot)$ denotes the mapping of the calibration for the LS estimator, $\mathcal{F}(\cdot)$ denotes the mapping of the calibration for ZF beamformer, and $\mathcal{H}(\cdot)$ denotes the uplink-to-downlink channel mapping \cite{Alrabeiah19FDD}. 

Since the mapping $\mathcal{H}(\cdot)$ from the uplink channel to the downlink channel has been extensively studied in the literature \cite{Alrabeiah19FDD} and the mapping $\mathcal{F}(\cdot)$ has been studied in the previous subsection, we focus on the mapping $\mathcal{B}(\cdot)$ here. As discussed in Section \uppercase\expandafter{\romannumeral2}, to guarantee the uplink antenna-wise permutation equivariance, we develop $M$ duplicate MLPs to process each row of $\mathbf{Y}_{\mathrm{p}}$. The structure shown in Fig. \ref{FrameworkStructure}(a) is utilized to approximate the mapping $\mathcal{B}(\cdot)$.


Due to the fast inference of deep neural networks and the high computational efficiencies of time-efficient methods, the proposed neural calibration-based end-to-end design achieves real-time implementation. Furthermore, since the conventional linear estimator and linear beamformer are asymptotically optimal for channel estimation and beamforming, respectively, they are scalable for large-scale wireless networks in terms of the achievable performance. In this way, by exploiting the domain knowledge inherent in these time-efficient methods and incorporating the intrinsic equivariance property introduced by the wireless network topology, the proposed neural calibration-based end-to-end design is more scalable and generalizable than the fully data-driven end-to-end method \cite{Sohrabi21FDD}. This will be verified via simulations in the next section.

\subsection{Training Strategy}
Since the existing deep learning platforms do not support complex-value operations, we rewrite the matrix multiplication $\mathbf{C} = \mathbf{A} \mathbf{B}$ as
\begin{equation}
\left[\begin{array}{cc}
\Re\left\{\mathbf{C}\right\} \\
\Im\left\{\mathbf{C}\right\}
\end{array}\right] = \left[\begin{array}{cc}
\Re\left\{\mathbf{A}\right\} & -\Im\left\{\mathbf{A}\right\} \\
\Im\left\{\mathbf{A}\right\} & \Re\left\{\mathbf{A}\right\}
\end{array}\right]
\left[\begin{array}{cc}
\Re\left\{\mathbf{B}\right\} \\
\Im\left\{\mathbf{B}\right\}
\end{array}\right].
\end{equation}
Besides, the ZF beamformer involves the matrix inverse $\left(\mathbf{H}_{\mathrm{DL}} \mathbf{H}_{\mathrm{DL}}^{H}\right)^{-1}$. Let $\mathbf{D}$ denote $\mathbf{H}_{\mathrm{DL}} \mathbf{H}_{\mathrm{DL}}^{H}$ and $\mathbf{E} = \mathbf{D}^{-1}$. Since $\mathbf{D} \mathbf{E} = \mathbf{I}$, the following equation is satisfied
\begin{equation}
\left[\begin{array}{cc}
\mathbf{I} \\
\mathbf{0}
\end{array}\right] = \left[\begin{array}{cc}
\Re\left\{\mathbf{D}\right\} & -\Im\left\{\mathbf{D}\right\} \\
\Im\left\{\mathbf{D}\right\} & \Re\left\{\mathbf{D}\right\}
\end{array}\right]
\left[\begin{array}{cc}
\Re\left\{\mathbf{E}\right\} \\
\Im\left\{\mathbf{E}\right\}
\end{array}\right].
\end{equation}
Therefore, $\Im\left\{\mathbf{E}\right\} = -\Re\left\{\mathbf{D}\right\}^{-1}\Im\left\{\mathbf{D}\right\}\Re\left\{\mathbf{E}\right\}$ and $\Re\left\{\mathbf{E}\right\} = \left(\Re\left\{\mathbf{D}\right\} + \Im\left\{\mathbf{D}\right\}\Re\left\{\mathbf{D}\right\}^{-1}\Im\left\{\mathbf{D}\right\}\right)^{-1}$. 

In the training stage, the negative objective functions in Problems \eqref{BF_perfectCSI} and \eqref{BF_implicit} are then readily to be used as the loss functions for the two applications, respectively. 
The trainable parameters are adjusted to minimize the loss function by employing the stochastic gradient descent (SGD) algorithm where the expectation in \eqref{BF_implicit} is approximated by the empirical average over the training samples \cite{Sohrabi21FDD}. The back propoagation procedures can be automatically implemented by any standard deep learning platform. Note that for the beamforming design with implicit channel estimation, the downlink CSI is only used during training to compute the loss function, and once trained, the operation of the neural network does not require any knowlege about downlink CSI. We also remark that the computationally-demanding training process is done offline, so it does not affect the computational complexity of the proposed method in the online inference stage.

%

\section{Simulation Results}
In this section, we demonstrate the performance of the proposed neural calibration-based resource allocation in massive MIMO systems.
\subsection{Simulation Setup}
In the experiments, we consider the indoor massive MIMO scenario ``I1'' that is provided in the DeepMIMO dataset \cite{Alkhateeb2019} and constructed based on the 3D ray tracing simulator \cite{Remcom}. 
Two operating frequencies, i.e., 2.4 GHz and 2.5 GHz, are employed as the uplink and downlink carrier frequencies, respectively.
We consider the ULA transmit antennas at the BS with the antenna spacing set at half wavelength and the number of propagation paths set to 5. 
The uplink pilot length is the same as the number of users and the noise power is set to $\sigma_0^2=\sigma_1^2=-85$ dBm in the simulations.
The user-wise shared MLP for ZF calibration has 4 fully-connected layers with 512, 2048, 2048, and $2M$ neurons in each layer, respectively, while the antenna-wise shared MLP for LS calibration has 4 fully-connected layers with 128, 512, 512, and $2L$ neurons in each layer, respectively. 
We train the neural network for 200 epochs using the Adam optimizer \cite{kingma2014adam} with a mini-batch size of 1024 and a learning rate of 0.001. There are in total 204,800 training samples and 1000 test samples. After each dense layer, the batch normalization layer is leveraged to accelerate convergence and the rectified linear unit (ReLU) is utilized as the activation function in the hidden layers. The numerical results are simulated on a desktop Intel (i7-8700) CPU running at 3.20 GHz with 16 GB RAM.

\subsection{Downlink Beamforming with Perfect Downlink CSI} 
To illustrate the effectiveness of the proposed neural calibration framework for beamforming design with perfect downlink CSI, we adopt five benchmarks for comparison:
\begin{itemize}
\item \textbf{WMMSE}: The conventional iterative WMMSE method in \cite{Shi11WMMSE} is used for beamforming. This baseline serves as a performance upper bound with a high computational complexity.
\item \textbf{ZF}: The ZF beamforming solution is employed.
\item \textbf{MRT}: The MRT beamforming solution is employed.
\item \textbf{Fully Data-driven}: The black-box CNN method in \cite{Hu21DeepUnfold} is used.
\item \textbf{Deep Unfolding}: The deep unfolding method in \cite{Hu21DeepUnfold} is utilized. To achieve satisfactory performance, $20$ iterations are used.
\end{itemize}

\begin{figure}[t] 
\centering
\subfigure[$K=8$ and $P_{\mathrm{DL}}=5$ dBm.]{
\begin{minipage}[t]{0.5\linewidth}
\centering
\includegraphics[width=3.4in]{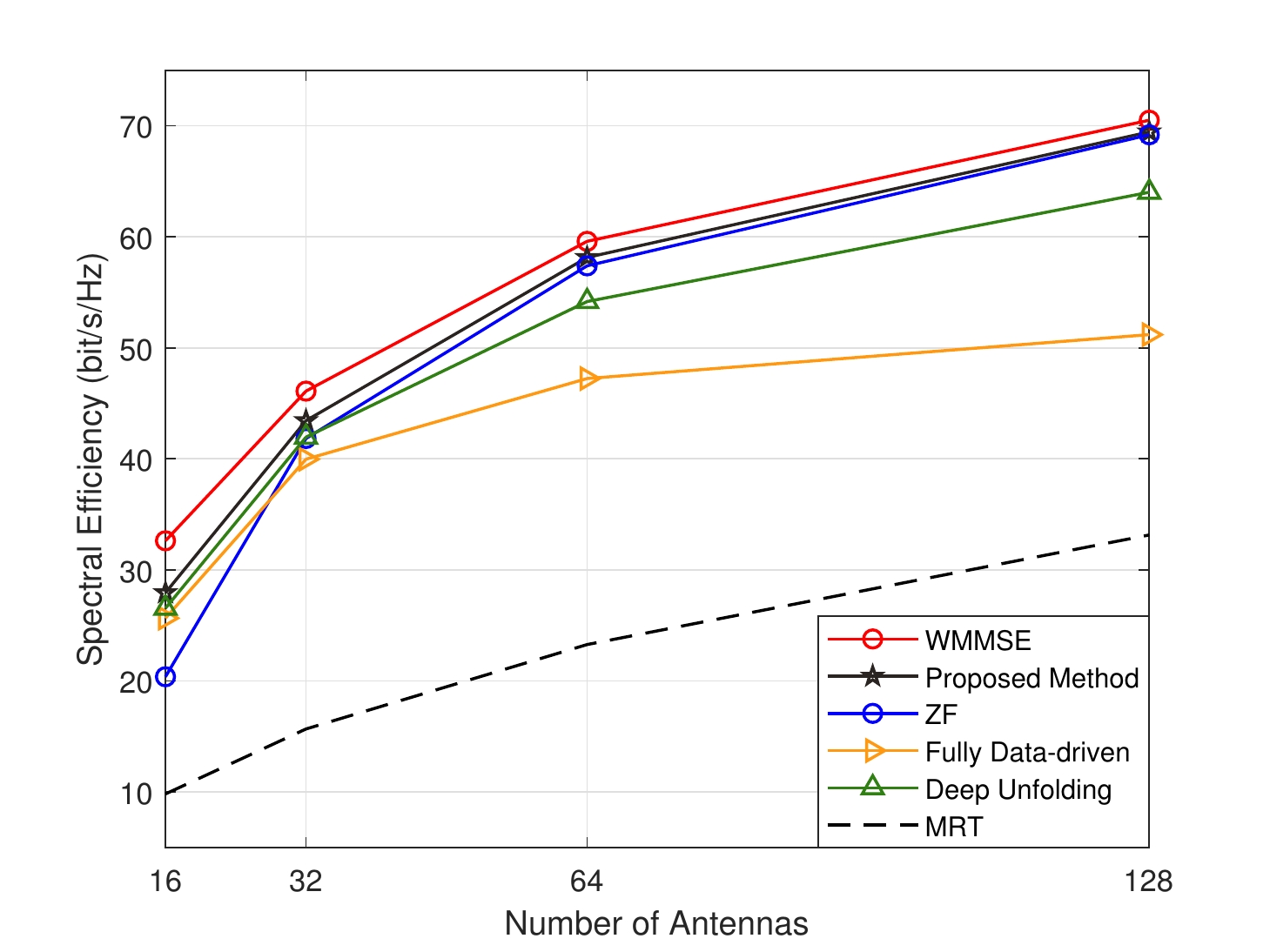}
\end{minipage}%
}%
\subfigure[$M=32$ and $P_{\mathrm{DL}}=5$ dBm.]{
\begin{minipage}[t]{0.5\linewidth}
\centering
\includegraphics[width=3.4in]{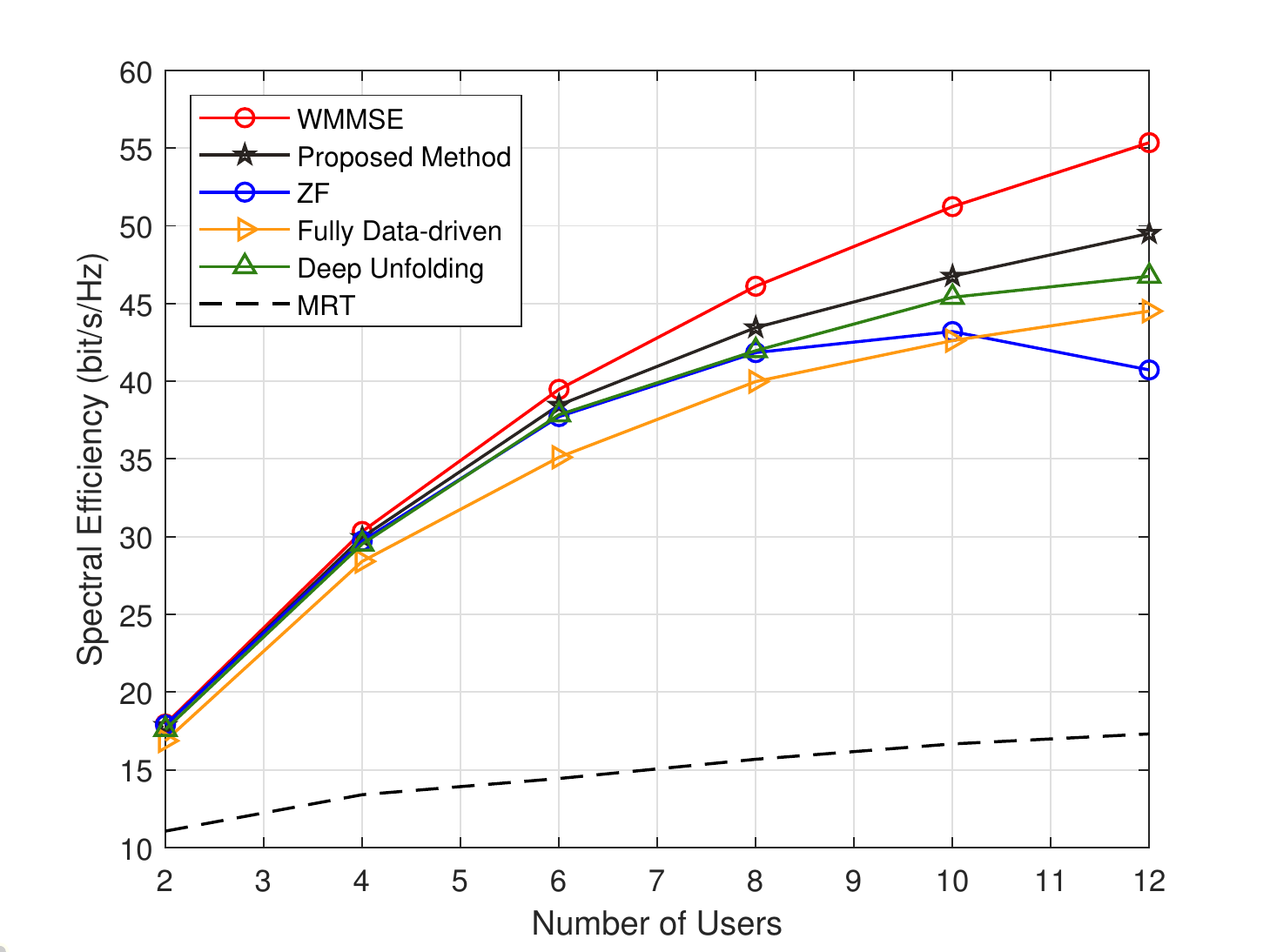}
\end{minipage}%
}%
\centering
\caption{Spectral efficiency achieved by different methods with perfect downlink CSI.}
\label{PerformancePerfect}
\end{figure}

Fig. \ref{PerformancePerfect}(a) plots the spectral efficiency achieved by the proposed scheme and the five baseline methods versus the number of antennas. It is shown that all the learning-based methods outperform the MRT method, indicating that deep learning approaches have the ability to effectively manage the interference. Among the three learning-based methods, the proposed neural calibration-based downlink beamforming scheme achieves the best performance. It is also observed that the proposed neural calibration-based design outperforms the conventional ZF beamformer for all investigated values of antenna size, and the performance gain is more obvious when the antenna number is small. This indicates that, when the ratio between the antenna number and the user number is small, i.e., when the ZF beamformer is far from optimal, by calibrating the input of the ZF beamformer using neural networks, the spectral efficiency can be effectively improved.
Furthermore, it is demonstrated in Fig. \ref{PerformancePerfect}(a) that the performance of our proposed method captures the trend of the WMMSE algorithm. In contrast, the fully data-driven method suffers from a significant performance degradation when $M$ increases. This clearly suggests that, by integrating domain knowledge into the beamforming design, our approach achieves a higher scalability compared to the fully data-driven method.

In Fig. \ref{PerformancePerfect}(b), we demonstrate the system spectral efficiency versus the number of users. As can be observed in Fig. \ref{PerformancePerfect}(b), while the ZF beamformer entails a prominent performance loss when the number of users increases, our proposed neural calibration-based method still captures the trend of the upper bound. This indicates that the calibration scheme is effectively learned from data thanks to the powerful learning capability of MLP.
Among all the three learning-based methods, the proposed method achieves the best performance for all investigated values of the user number.
This verifies the superiority of the proposed neural calibration design in terms of both the spectral efficiency and scalability in wireless networks where users are densely distributed.

\begin{figure}[t] 
\centering
\includegraphics[width=3.4in]{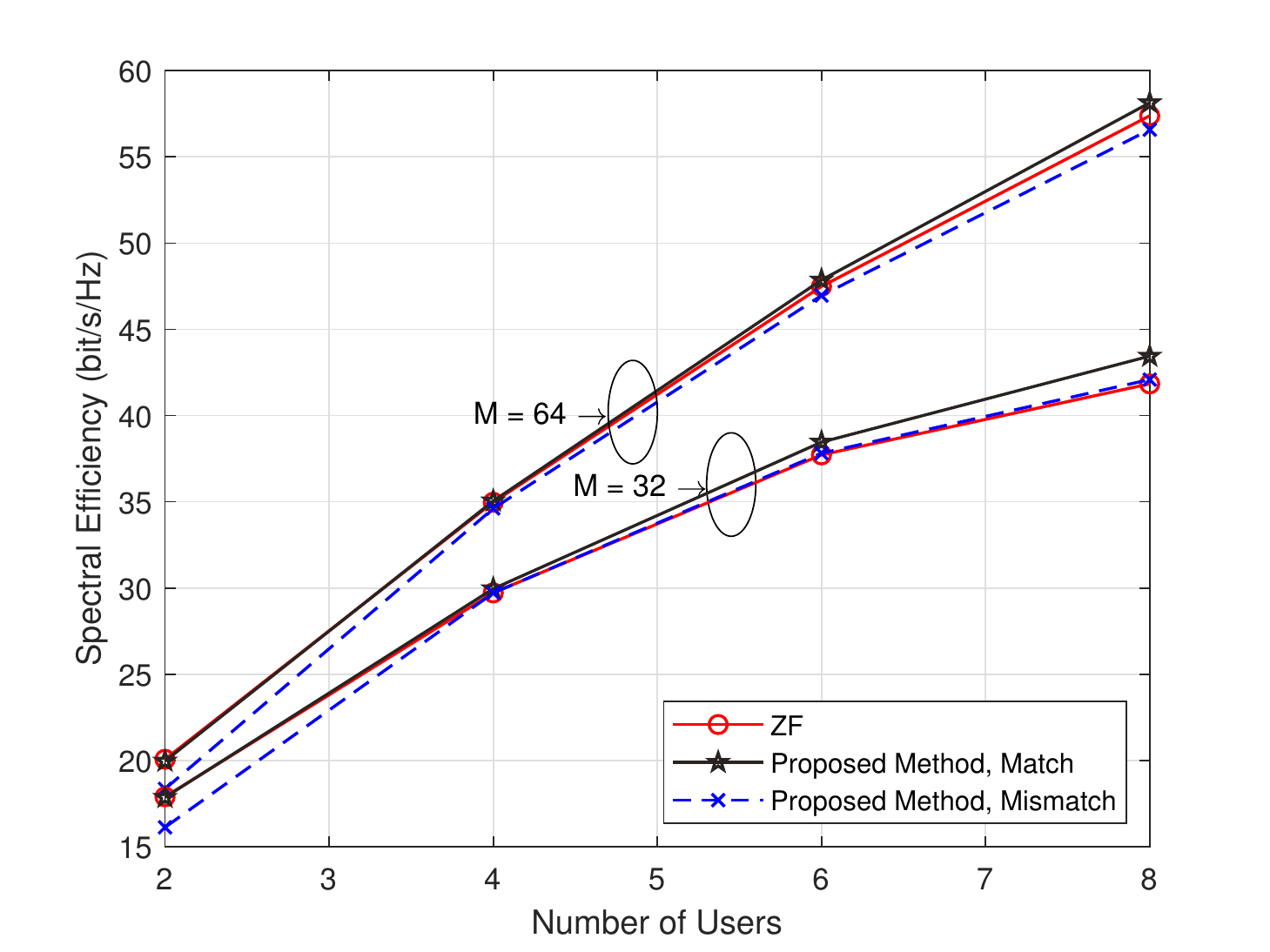} 
\caption{The task-mismatch generalization of the proposed method when $P_{\mathrm{DL}}=5$ dBm.} 
\label{figure00} 
\end{figure}

We also test the generalization ability of our method in terms of the variable number of users and for different number of antennas, $M$, in Fig. \ref{figure00}. Since the input and output dimensions of the proposed neural network are independent of the user number, the deep learning model trained for a certain value of $K$ can be directly implemented when the user number varies. The ``Proposed Method, Mismatch” indicates the case where the learning model is trained when $K=4$ but tested for different number of users, while in ``Proposed Method, Match”, the training setting is exactly the same as the test setting. 
It is shown that there is little performance loss in terms of the spectral efficiency when mismatch exists in terms of the number of users. Specifically, when the number of antennas is 64, the performance of the mismatched learning model is at least $93.24\%$ of that achieved by the perfectly matched learning model for all investigated values of user number. This confirms the high generalization ability of the proposed neural calibration design.

\begin{figure}[t] 
\centering
\includegraphics[width=3.4in]{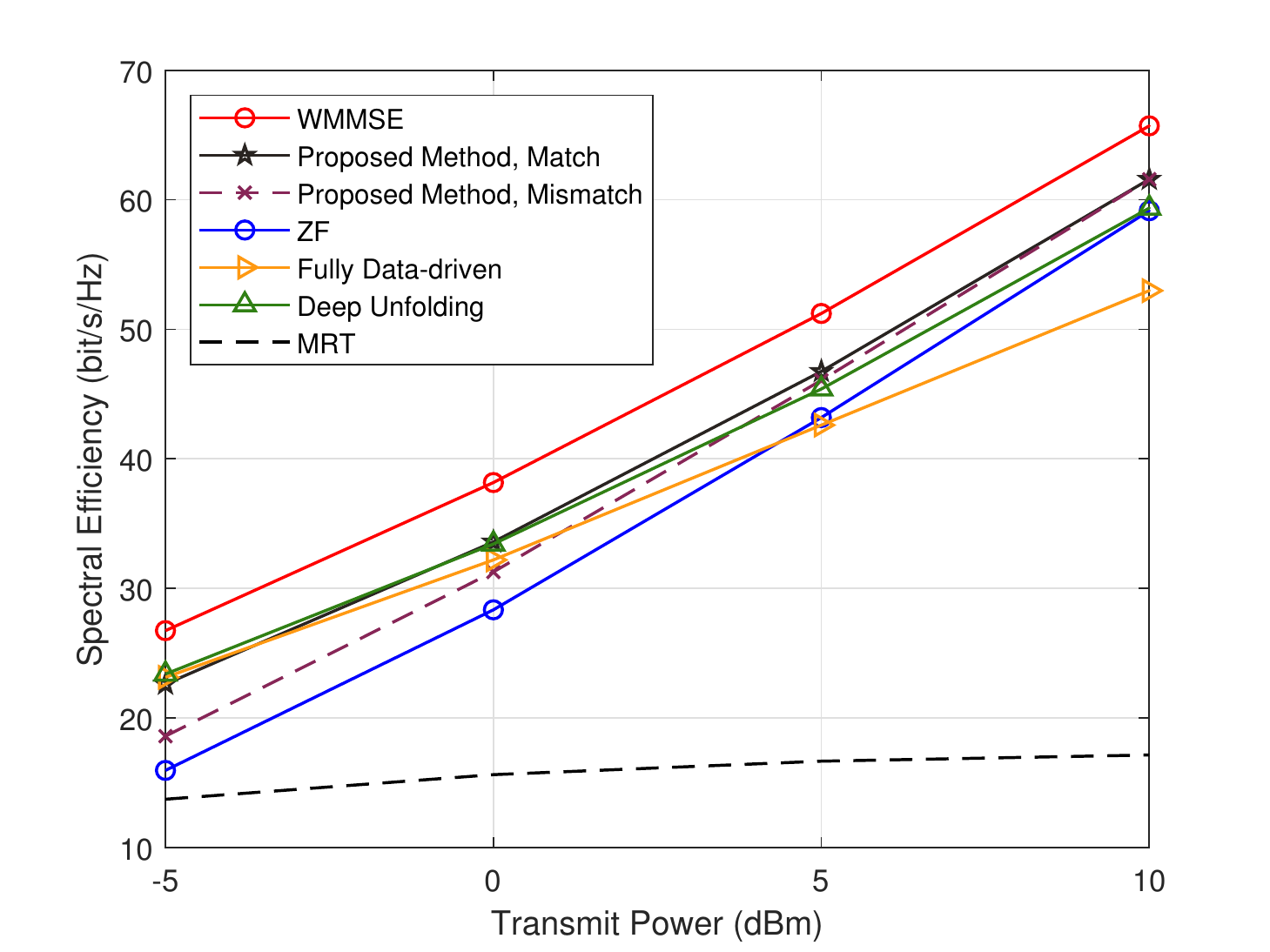} 
\caption{Spectral efficiency achieved by different methods when the number of antennas $M=32$ and the number of users $K=10$.} 
\label{SNR} 
\end{figure}

We then test the performance of our method versus different downlink transmit powers. The ``Proposed Method, Mismatch” indicates the case where the learning model is trained when $P_{\mathrm{DL}}=-10$ dBm but tested for different downlink transmit powers, while in ``Proposed Method, Match”, the training setting is exactly the same as the test setting. In Fig. \ref{SNR}, the spectral efficiency achieved by the matched neural calibration method is significantly higher than that of the ZF beamformer over the whole SNR regime. The fully data-driven end-to-end benchmark achieves good performance in the low SNR regime, but when the downlink transmit power increases, it fails to outperform the conventional ZF beamformer. This indicates that when the power of interference signals is strong, fully data-driven method fails to effectively manage the interference, which leads to a significant performance loss. When mismatch exists, the proposed learning model only results in little performance loss and still outperforms the ZF beamformer for all investigated values of downlink transmit power. This shows that the proposed neural calibration-based beamforming design generalizes well under different downlink SNR.

\begin{table*}[t]	
\selectfont  
\centering
\newcommand{\tabincell}[2]{\begin{tabular}{@{}#1@{}}#2\end{tabular}}
\caption{The running time of different methods under different settings (in millisecond).} 
\begin{tabular}{|c|c|c|c|c|}
\hline
($M$, $K$) & WMMSE & Neural Calibration & Deep Unfolding & Fully Data-driven \\ \hline
(16, 2) & 32.06 & 0.14 & 16.97 & 0.11  \\ \hline
(32, 2) & 52.99 & 0.15 & 22.11 & 0.14   \\ \hline
(32, 8) & 171.36 &  0.46 & 152.92 & 0.43   \\ \hline
(32, 12) & 291.99 &  0.70 & 297.89 & 0.32  \\ \hline
(64, 4) & 133.20 & 0.26 & 103.86 & 0.28 \\ \hline
(64, 12) & 415.03 &  0.71 & 381.89 & 0.58  \\ \hline
\end{tabular}
\label{run-time}
\end{table*}

We then show the average running time of different methods in Table \ref{run-time}. Due to the matrix inverse and bisection search operation in each iteration, the WMMSE method entails the longest running time, which defers their application in practice. Since the deep unfolding approach is based on the WMMSE algorithm, it is still an iterative method and fails to be executed at the millisecond level. It can be observed that the proposed neural calibration method and the fully data-driven method both enjoy a high computational efficiency and can achieve real-time implementation.

\subsection{Downlink Beamforming with Implicit Channel Estimation}
To illustrate the effectiveness of the proposed neural calibration end-to-end design for \eqref{BF_implicit}, we adopt four benchmarks for comparisons:
\begin{itemize}
\item \textbf{WMMSE Perfect CSIT}: Assuming perfect downlink CSI at the transmitter (CSIT), the conventional iterative WMMSE method in \cite{Shi11WMMSE} is used for beamforming. This baseline serves as a performance upper bound with a high computational complexity.
\item \textbf{ZF Perfect CSIT}: The ZF beamforming solution is employed given perfect CSIT.
\item \textbf{Fully Data-driven End-to-End}: The black-box CNN method in \cite{Hu21DeepUnfold} is modified to map the received uplink pilot to the downlink precoding matrix.
\item \textbf{LS + Channel Mapping + ZF}: Conventional LS channel estimator, channel mapping in \cite{Alrabeiah19FDD}, and ZF beamformer are adopted in a block-by-block manner.
\end{itemize}

\begin{figure}[t] 
\centering
\subfigure[$K=8$, $P_{\mathrm{UL}}=-10$ dBm, and $P_{\mathrm{DL}}=5$ dBm.]{
\begin{minipage}[t]{0.5\linewidth}
\centering
\includegraphics[width=3.4in]{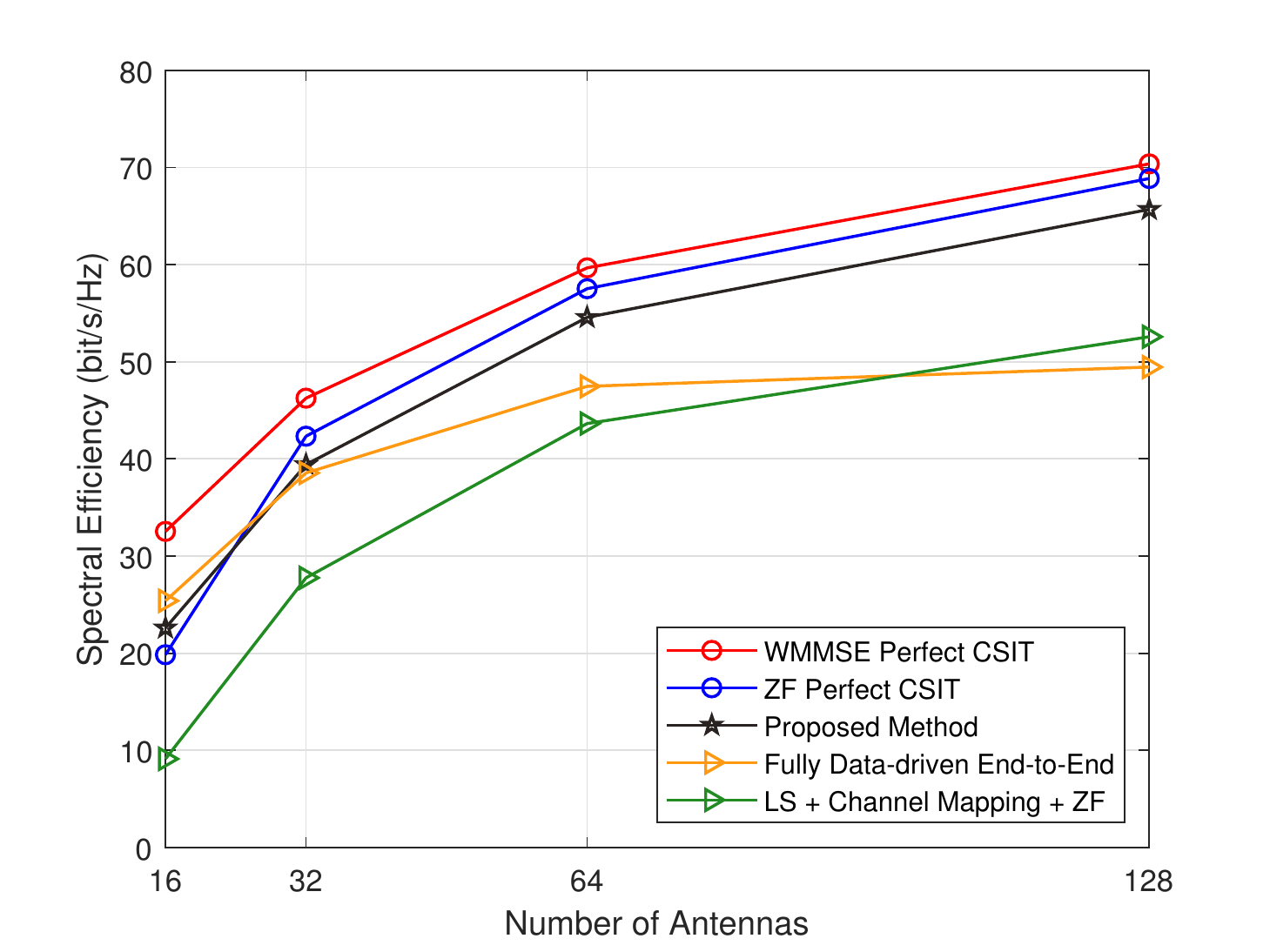}
\end{minipage}%
}%
\subfigure[$M=64$, $P_{\mathrm{UL}}=-10$ dBm, and $P_{\mathrm{DL}}=5$ dBm.]{
\begin{minipage}[t]{0.5\linewidth}
\centering
\includegraphics[width=3.4in]{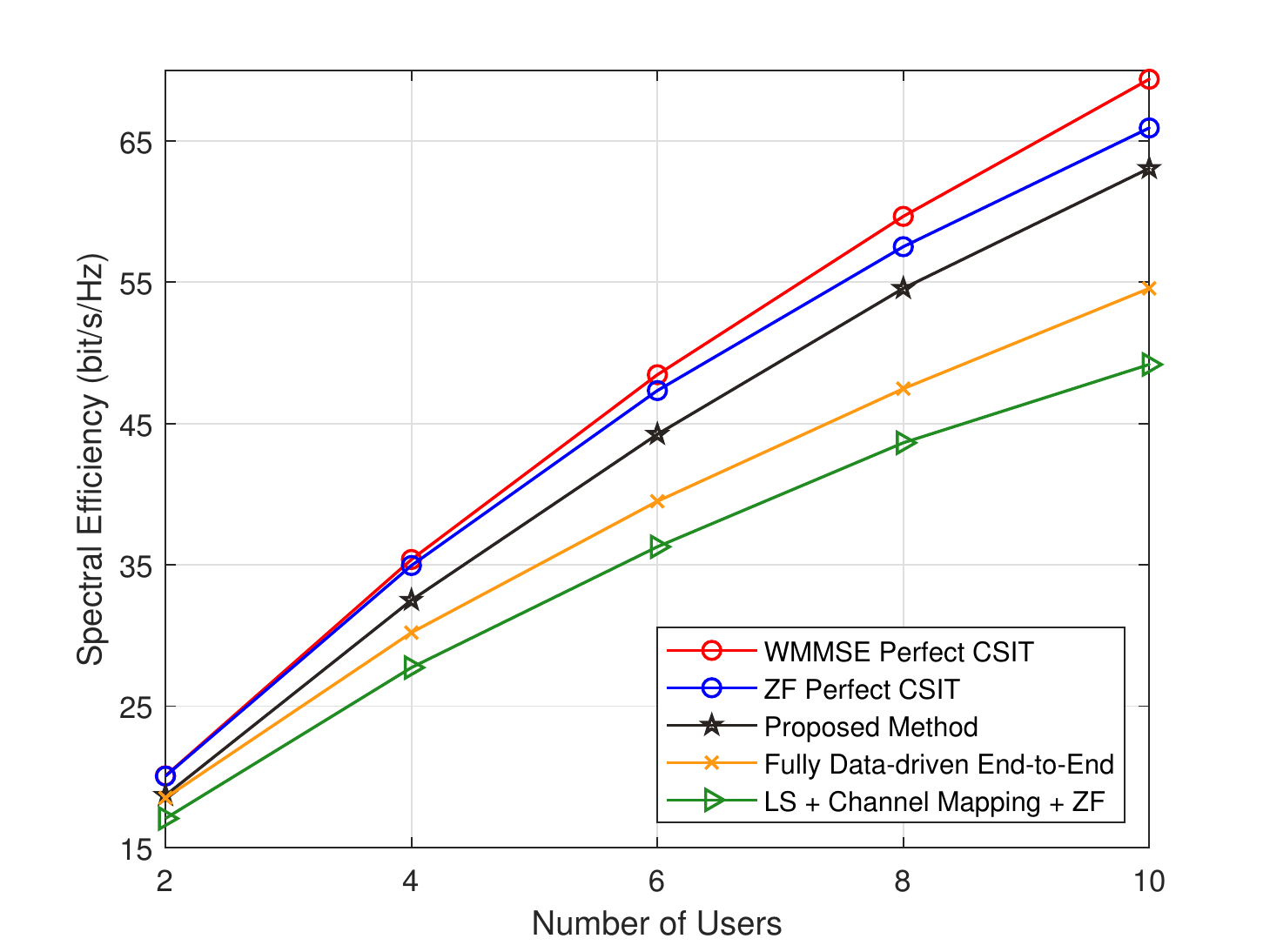}
\end{minipage}%
}%
\centering
\caption{Spectral efficiency achieved by different methods.}
\label{PerformanceImplicit}
\end{figure}

%
Fig. \ref{PerformanceImplicit}(a) plots the spectral efficiency achieved by the proposed scheme and the four baseline methods versus the number of antennas. It can be observed that the proposed neural calibration-based design significantly outperforms the conventional block-by-block method for all investigated values of the antenna size.
Besides, the proposed method even outperforms ZF with perfect CSIT when the number of antennas $M$ is small. This shows that, by calibrating the input of the ZF beamformer using neural networks, the system performance in the small-scale antenna regime can be effectively improved. Furthermore, it is demonstrated in Fig. \ref{PerformanceImplicit}(a) that the performance of our proposed method captures the trend of the two baseline beamformers with perfect CSIT. In contrast, the fully data-driven end-to-end method suffers from a significant performance degradation when $M$ increases, and its performance gain over the block-by-block method vanishes when $M = 128$. This clearly indicates that, by integrating domain knowledge into the end-to-end design, our approach achieves a higher scalability compared to the fully data-driven end-to-end method.

In Fig. \ref{PerformanceImplicit}(b), we demonstrate the system spectral efficiency versus the number of users. As can be observed in Fig. \ref{PerformanceImplicit}(b), while the two baseline methods without perfect CSIT entail a prominent performance loss when the number of users increases, our proposed neural calibration-based method only experiences little performance loss.
Specifically, the proposed beamforming method can achieve $91.0\%$ of the spectral efficiency achieved by the WMMSE with perfect CSIT when $K=10$, while the fully data-driven end-to-end method only achieves $78.7\%$ in this case. This verifies the superiority of the proposed neural calibration design in terms of both the spectral efficiency and scalability in wireless networks where users are densely distributed.

\begin{figure}[t] 
\centering
\includegraphics[width=3.4in]{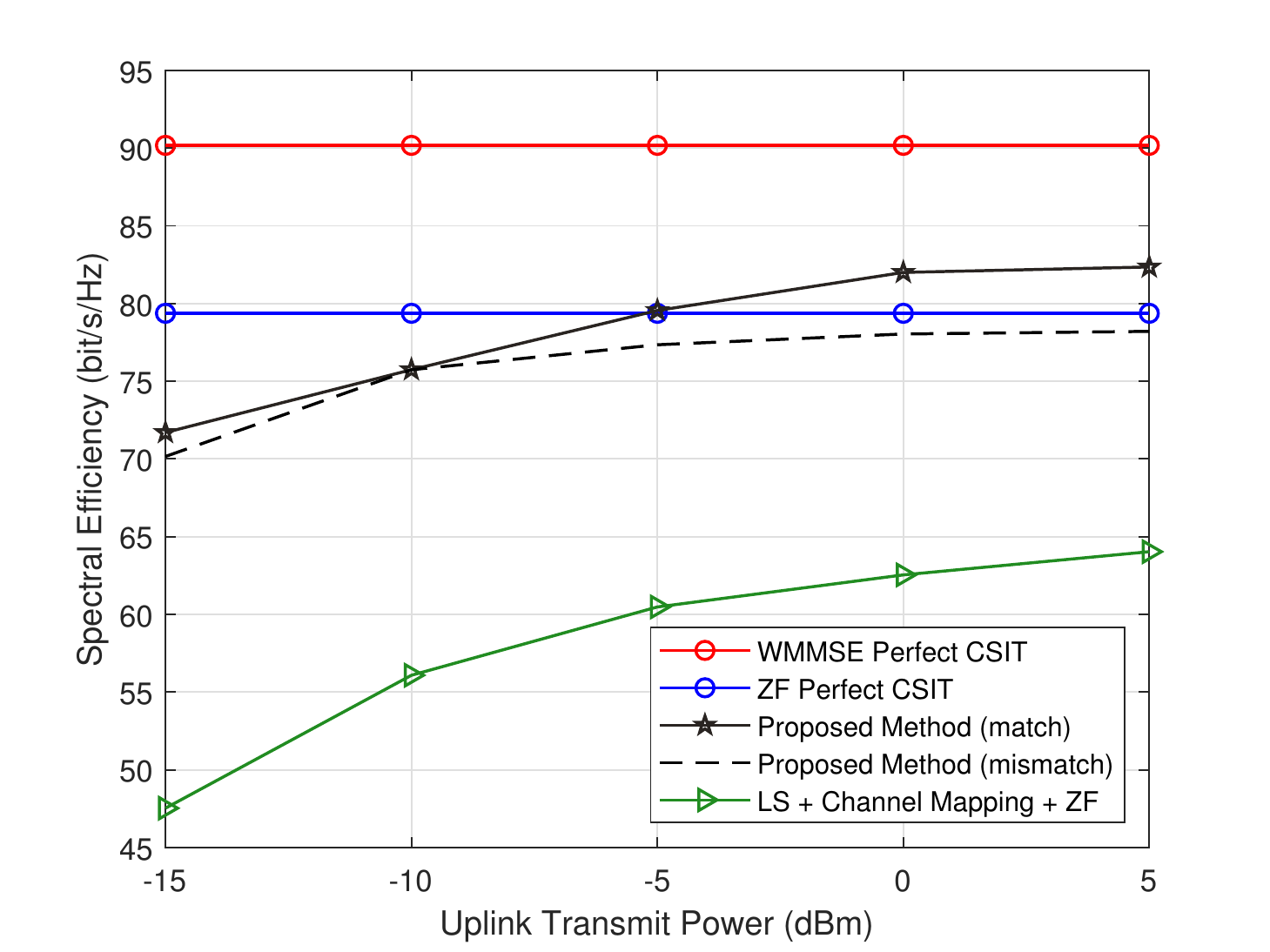} 
\caption{Spectral efficiency achieved by different values of the uplink transmit powers when $M=64$, $K=16$, and $P_{\rm{DL}}=5$ dBm.} 
\label{figure3} 
\end{figure}
We also test the robustness of our method against mismatch in terms of the uplink SNR. In Fig. \ref{figure3}, the ``Proposed Method (mismatch)” indicates the case where the learning model is trained when $P_{\rm{UL}}=-10$ dBm but tested for different values of the uplink transmit power. 
It is shown that there is little performance loss in terms of the spectral efficiency when mismatch exists for the uplink transmit power, which confirms the robustness of the proposed neural calibration end-to-end design. Besides, the spectral efficiency achieved by both the matched and mismatched neural calibrations is significantly higher than that of the block-by-block design over the whole SNR regime, which indicates the superiority of the neural calibration-based resource allocation with implicit channel estimation. Since the fully data-driven end-to-end benchmark is not scalable for dense wireless networks, its performance is not shown here.

\section{Conclusions}
In this paper, we developed a scalable and generalizable neural calibration framework to solve generic design problems in wireless communication systems. In contrast to two existing ``learning to communicate” paradigms, i.e., fully data-driven methods and deep unfolding methods, we proposed a novel way to amalgamate domain knowledge with deep learning. The backbone of the conventional time-efficient method is retained and a unique neural network architecture is developed based on a permutation equivariance property introduced by the wireless network topology. Notable advantages of the proposed method include real-time implementation, scalable performance for large-scale problems, and good generalizations to varying system parameters. Simulation results clearly demonstrated that the proposed neural calibration-based design achieves an excellent performance in large-scale massive MIMO systems even if perfect CSI is not available. It is intriguing to extend our proposed neural calibration framework to other challenging communication applications, such as system design with pilot contamination \cite{Shen15Downlink}, beamforming design in multicell systems \cite{Jun09Precoding}, and design problems in intelligent reflecting surfaces-aided wireless systems \cite{yu2021smart}.

\appendices 
\section{Proof of Theorem \ref{UplinkPE}} \label{proof:UplinkPE}
\begin{IEEEproof}
Reordering the antennas simply leads to permutation of the rows of $\mathbf{H}$ and ${\mathbf{Y}_\mathrm{p}}$. The antenna ordering permutation equivariance property of $\mathbf{H}$ and ${\mathbf{Y}_\mathrm{p}}$ is depicted as
\begin{equation}
{\mathbf{Y}}_\mathrm{p}^{'} = \mathbf{\Pi}^T {\mathbf{Y}_\mathrm{p}} = \mathbf{\Pi}^T \mathbf{H} {\mathbf{P}} + {\mathbf{N}} = \mathbf{H}^{'} {\mathbf{P}} + {\mathbf{N}}, \quad \forall \mathbf{\Pi}.
\end{equation}
This suggests that there is no natural ordering of antenna elements. Let $\hat{\mathbf{H}} = \mathcal{P}({\mathbf{Y}_\mathrm{p}})$ and $\hat{\mathbf{H}}^{'} = \mathcal{P}({\mathbf{Y}}_\mathrm{p}^{'})$ denote the estimates of the channel matrix given ${\mathbf{Y}_\mathrm{p}}$ and ${\mathbf{Y}}_\mathrm{p}^{'}$, respectively. 
Since the estimated channels should follow the same permutation of the true channels, i.e., $\hat{\mathbf{H}}^{'} = \mathbf{\Pi}^T \hat{\mathbf{H}}$, the permutation equivariance property of the channel estimator is stated as
\begin{equation}
\mathcal{P}(\mathbf{\Pi}^T {\mathbf{Y}_\mathrm{p}}) = \mathbf{\Pi}^T \mathcal{P}({\mathbf{Y}_\mathrm{p}}).
\end{equation}
The proof is thus completed.
\end{IEEEproof}

\section{Proof of Theorem \ref{Potential}} \label{proof:Potential}
\begin{IEEEproof}
The gradient of the sum-rate function with respect to $\mathbf{V}$ is ${\nabla _{\mathbf{V}}}R = {\mathbf{H}}_{\mathrm{DL}}^H{\mathbf{B}}$ where $\mathbf{B} \in \mathbb{C}^{K \times K}$ and its entries are given by
\begin{equation}
\begin{aligned}
{b_{kk}} = ~& \frac{{\mathbf{h}_{\mathrm{DL},k}^H{{\mathbf{v}}_k}}}{{\sum\limits_{i = 1}^K {|\mathbf{h}_{\mathrm{DL},k}^H{{\mathbf{v}}_i}{|^2} + {\sigma_0 ^2}} }}, \\
{b_{jk}} = ~& \frac{{ - |\mathbf{h}_{\mathrm{DL},j}^H{{\mathbf{v}}_j}{|^2}\mathbf{h}_{\mathrm{DL},j}^H{{\mathbf{v}}_k}}}{{(\sum\limits_{i = 1}^K {|\mathbf{h}_{\mathrm{DL},j}^H{{\mathbf{v}}_i}{|^2} + {\sigma_0 ^2}} )(\sum\limits_{i \ne j} {|\mathbf{h}_{\mathrm{DL},j}^H{{\mathbf{v}}_i}{|^2} + {\sigma_0 ^2}} )}}, j \neq k.
\end{aligned}
\end{equation}
Since $\mathbf{V}_{\mathrm{ZF}} = h_{\mathrm{ZF}}(\mathbf{X})$, the gradient of the sum-rate function with respect to the input of ZF, i.e., $\mathbf{X}$, is given by
\begin{equation}
\begin{aligned}
{\nabla _{\mathbf{X}}}R = {\gamma_{\mathrm{ZF}}}[ ~& {(\mathbf{X}{\mathbf{X}^H})^{ - 1}} \mathbf{B}^{H} \mathbf{H}_{\mathrm{DL}}  - \mathbf{X}^ {'} \mathbf{H}_{\mathrm{DL}}^{H} \mathbf{B} \mathbf{X}^ {'} \\
~& - {(\mathbf{X}{\mathbf{X}^H})^{ - 1}} \mathbf{B}^{H} \mathbf{H}_{\mathrm{DL}} {\mathbf{X}^{H}} \mathbf{X}^ {'} ],
\end{aligned}
\end{equation}
where $\mathbf{X}^ {'} = {({\mathbf{X}}{{\mathbf{X}}^H})^{ - 1}} {\mathbf{X}}$.
It can be observed that when $\sigma_0^2 \neq 0$, the gradient of the sum-rate function with respect to $\mathbf{X}$ is not zero at $\mathbf{X}=\mathbf{H}_{\mathrm{DL}}$.
Therefore, when $\sigma_0^2 \neq 0$, $\forall \mathbf{H}_{\mathrm{DL}}$ and $\forall \epsilon > 0$, there exists an $\widetilde{\mathbf{H}}_{\mathrm{DL}}$ with $\|\mathbf{H}_{\mathrm{DL}} - \widetilde{\mathbf{H}}_{\mathrm{DL}}\| \leq \epsilon$ such that $R\left(h_{\mathrm{ZF}}(\mathbf{H}_{\mathrm{DL}})\right) < R(h_{\mathrm{ZF}}(\widetilde{\mathbf{H}}_{\mathrm{DL}}))$. 
\end{IEEEproof}

\section{Proof of Theorem \ref{perm}} \label{proof:perm}
\begin{IEEEproof}
First, since $\mathcal{F}(\cdot)$ consists of one forward pass of the MLP $\phi$ and the stack operation, it natrually satisfies the permutation equivariance. It is easy to verify that $\mathcal{F}(\mathbf{\Pi}^T \mathbf{H}_{\mathrm{DL}}) = \mathbf{\Pi}^T \mathcal{F}( \mathbf{H}_{\mathrm{DL}}) = \mathbf{\Pi}^T \widetilde{\mathbf{H}}_{\mathrm{DL}}$.  Then, to show $h_{\mathrm{ZF}}(\mathcal{F}(\mathbf{\Pi}^T \mathbf{H}_{\mathrm{DL}})) = h_{\mathrm{ZF}}(\mathcal{F}(\mathbf{H}_{\mathrm{DL}})) \mathbf{\Pi}$, we need to prove that $h_{\mathrm{ZF}}(\mathbf{\Pi}^T \widetilde{\mathbf{H}}_{\mathrm{DL}}) = h_{\mathrm{ZF}}(\widetilde{\mathbf{H}}_{\mathrm{DL}}))\mathbf{\Pi}$.

Since $\mathbf{\Pi} \mathbf{\Pi}^T = \mathbf{I}$, then for any invertible matrix $\mathbf{A}$, we have $\left(\mathbf{\Pi} \mathbf{A} \mathbf{\Pi}^T \right) ^ {-1} = \mathbf{\Pi} \mathbf{A}^{-1} \mathbf{\Pi}^T$. Therefore,
\begin{equation}
\begin{aligned}
h_{\mathrm{ZF}}(\mathbf{\Pi}^T \widetilde{\mathbf{H}}_{\mathrm{DL}}) & = \gamma_{\mathrm{ZF}} (\mathbf{\Pi}^T \widetilde{\mathbf{H}}_{\mathrm{DL}})^H \left(\mathbf{\Pi}^T \widetilde{\mathbf{H}}_{\mathrm{DL}} \left(\mathbf{\Pi}^T \widetilde{\mathbf{H}}_{\mathrm{DL}}\right)^H \right)^{-1} \\
& = \gamma_{\mathrm{ZF}} (\widetilde{\mathbf{H}}_{\mathrm{DL}})^H \mathbf{\Pi} \left(\mathbf{\Pi}^T \widetilde{\mathbf{H}}_{\mathrm{DL}} \widetilde{\mathbf{H}}_{\mathrm{DL}}^H \mathbf{\Pi} \right)^{-1} \\
& = \gamma_{\mathrm{ZF}} \widetilde{\mathbf{H}}_{\mathrm{DL}}^H \mathbf{\Pi} \mathbf{\Pi}^T \left(\widetilde{\mathbf{H}}_{\mathrm{DL}} \widetilde{\mathbf{H}}_{\mathrm{DL}}^H \right)^{-1} \mathbf{\Pi} \\
& = \gamma_{\mathrm{ZF}} \widetilde{\mathbf{H}}_{\mathrm{DL}}^H \left(\widetilde{\mathbf{H}}_{\mathrm{DL}} \widetilde{\mathbf{H}}_{\mathrm{DL}}^H \right)^{-1} \mathbf{\Pi}.
\end{aligned}
\end{equation}

The proof is thus completed.
\end{IEEEproof}

\bibliographystyle{IEEEtran}
\bibliography{IEEEabrv,learning}

\end{document}